\documentclass[aps,pre,preprint,onecolumn,citeautoscript,superscriptaddress]{revtex4-1}  
\usepackage{amsmath,amssymb,mathrsfs,bm,feynmf,setspace,xspace,soul} 
\usepackage{graphicx}  
\usepackage[tight]{subfigure}    
\usepackage{color} 
\usepackage[papersize={8.5in,11in}]{geometry}
\usepackage[colorlinks=true]{hyperref}  
\hypersetup{
    bookmarks=true,         
    unicode=false,          
    pdftoolbar=true,        
    pdfmenubar=true,        
    pdffitwindow=false,     
    pdfstartview={FitH},    
    pdftitle={My title},    
    pdfauthor={Author},     
    pdfsubject={Subject},   
    pdfcreator={Creator},   
    pdfproducer={Producer}, 
    pdfkeywords={keyword1} {key2} {key3}, 
    pdfnewwindow=true,      
    colorlinks=true,       
    linkcolor=magenta, 
    citecolor=blue,        
    filecolor=magenta,      
    urlcolor=cyan           
} 

\newcommand{\al}{\alpha} 
 
\newcommand{\g}{\gamma}
\newcommand{\de}{\delta} 
\newcommand{\e}{\epsilon} 
 
\newcommand{\z}{\zeta}

\newcommand{\s}{\sigma}

\newcommand{\w}{\omega}

\newcommand{\De}{\Delta} 
\newcommand{\G}{\Gamma}

\newcommand{\pd}{\partial}

\newcommand{\ang}[1]{\left\langle #1 \right\rangle}

\newcommand{\beq}{\begin{equation}}
\newcommand{\eeq}{\end{equation}}
\newcommand{\ba}{\begin{array}{ccc}}
\newcommand{\ea}{\end{array}}
\newcommand{\nn}{\nonumber \\}
\newcommand{\bx}{{\bm x}}
\newcommand{\bk}{{\bm k}}
\newcommand{\bp}{{\bm p}}

\newcommand{\bomega}{{\bm \omega}}
\def\bea{\begin{eqnarray}}
\def\eea{\end{eqnarray}}
\newcommand{\bml}{\begin{multline}}

\newcommand{\eeqm}{\end{multline}}

\newcommand{\bsp}{\begin{split}}
\newcommand{\esp}{\end{split}}

\renewcommand{\b}[1]{{\bm #1}}
\renewcommand{\t}{\tilde}

\newcommand{\mc}{\mathcal}

\renewcommand{\t}{\tilde}
\newcommand{\ra}{\rightarrow}
\newcommand{\req}[1]{Eq.\thinspace(\ref{eq:#1})} 

\newcommand{\rfig}[1]{Fig.~\ref{fig:#1}}

\newcommand{\ie}{{\em i.e.\/}\@\xspace} 
\newcommand{\ts}{\thinspace{}}

\DeclareMathOperator{\tr}{tr}

\DeclareMathOperator{\Li}{Li}
\DeclareMathOperator{\im}{Im}
\DeclareMathOperator{\re}{Re}

\newcommand{\enm}{energy-momentum }

\begin{document}

\title{Conformal field theories at non-zero temperature:\\ operator product expansions, Monte Carlo, and holography}  
\author{Emanuel Katz}
\affiliation{Physics Department, Boston University, 590 Commonwealth Ave., Boston, MA 02215, USA}
 \author{Subir Sachdev}
 \affiliation{Department of Physics, Harvard University, Cambridge, Massachusetts, 02138, USA}
 \affiliation{Perimeter Institute for Theoretical Physics, Waterloo, Ontario N2L 2Y5, Canada}
\author{Erik S. S{\o}rensen}
 \affiliation{Department of Physics \& Astronomy, McMaster University, Hamilton, Ontario L8S 4M1, Canada}
 \author{William Witczak-Krempa}
 \affiliation{Perimeter Institute for Theoretical Physics, Waterloo, Ontario N2L 2Y5, Canada}
 \date{\today\\
 \vspace{0.6in}}
\begin{abstract} 
We compute the non-zero temperature conductivity of conserved flavor currents in conformal field theories (CFTs) in 2+1 spacetime dimensions.
At frequencies much greater than the temperature, $\hbar \omega \gg k_B T$, the $\omega$ dependence can be computed from the operator
product expansion (OPE) between the currents and operators which acquire a non-zero expectation value at $T>0$. Such results are
found to be in excellent agreement with quantum Monte Carlo studies of the O(2) Wilson-Fisher CFT. Results for the conductivity
and other observables are also obtained in vector $1/N$ expansions. We match these large $\omega$ results to the corresponding correlators of holographic
representations of the CFT: the holographic approach then allows us to extrapolate to small $\hbar \omega/(k_B T)$. 
Other holographic studies implicitly only used the OPE between the currents and the energy-momentum tensor, and this yields
the correct leading large $\omega$ behavior for a large class of CFTs. However, for the Wilson-Fisher CFT a relevant ``thermal'' operator 
must also be considered, and then consistency with the Monte Carlo results is obtained without a previously needed ad hoc rescaling of the $T$ value.
We also establish sum rules obeyed by the conductivity of a wide class of CFTs.
\end{abstract}
\maketitle 
\tableofcontents  
\section{Introduction} 
\label{sec:intro}   
 
Conformal field theories (CFTs) constitute the best characterized quantum systems without quasiparticle excitations. 
Their non-zero temperature dissipative dynamics can be treated by extensions of Boltzmann-like approaches designed
for quasiparticle dynamics \cite{damle}; the Boltzmann approach is difficult in general, 
thus limited in practice.
Much additional insight can be gained from a modern perspective 
based upon holographic ideas \cite{krempa_nature}, which does not assume a quasiparticle decomposition of the spectrum at any stage.
CFTs are also important as models of quantum critical points in condensed matter, 
notably for the superfluid-insulator
transition of bosons in a periodic potential in two spatial dimensions \cite{spielman,zhang,endres}.

In recent work  by three of us \cite{krempa_nature},  we computed the $T>0$ conductivity of a lattice model 
for this superfluid-insulator transition
using quantum Monte Carlo simulations;
after carefully taking the $T \rightarrow 0$ limit of the lattice model, we obtained
the $T>0$ conductivity of a conserved current of the CFT, and this was compared 
with the predictions of a semi-phenomenological holographic theory. The latter theory included terms up to four derivatives in 
the metric and a gauge field conjugate to the conserved current. We found consistency between the two approaches
after an ad hoc rescaling of the temperature between the two methods. Related $T>0$ results were obtained in Refs.\ts\onlinecite{sorensen,pollet},
$T=0$ results are in Refs.\ts\onlinecite{gazit,gazit14}, and the effects of disorder were considered 
in Refs.\ts\onlinecite{Lin11,trivedi}.  

The present paper will significantly improve on our previous analysis by using more specific field-theoretic information on 
the CFTs under consideration. We will work mainly with the 2+1 dimensional CFT
with O($N$) symmetry described by the Wilson-Fisher fixed point, and determine the conductivity of the conserved O($N$) current.
We will compute the operator product expansion (OPE) of the current operators in terms of other operators of the CFT, and use this
to constrain the high frequency behavior of the conductivity. We find excellent agreement of such results with Monte Carlo
studies of the O(2) model upon taking into account a scalar field conjugate to a relevant perturbation of the CFT. 
Next, we will connect the high frequency behavior to holography, and use it 
to make predictions for the conductivity at lower frequencies {\em without\/} an ad hoc rescaling of temperature.

From a broader perspective, our analysis shows how the finite temperature properties of CFTs can be analyzed
by systematically including the influence of low dimension operators to constrain the short-time behavior, and then using holography
to extrapolate to longer times. In theories with quasiparticles, the extrapolation from short to long times is generally made via
the Boltzmann equation; here, we argue that the corresponding extrapolation for CFTs without quasiparticles can be made by
a combination of the OPE with holography. 

We present here the structure of the high frequency, or short time, behavior of the conductivity as given by the OPE for a general
CFT  in 2+1 dimensions.  
With spacetime co-ordinates $(\tau, x, y)$, the conductivity is related to the two-point correlator of a conserved current
$J_x$ (we suppress indices of global flavor symmetries). We work in the Euclidean time signature, and then the conductivity is
\beq
\frac{\sigma (i\omega_n)}{\sigma_Q} = - \frac{1}{\omega_n} \left\langle J_x (\bomega) J_x (- \bomega) \right\rangle_{T} + \mbox{a possible contact term}\,, \label{eq:condJJ}
\eeq
where $\bomega \equiv (\omega_n, 0, 0)$, and in some cases a diamagnetic ``contact'' term may be present (this
is the case for the O$(N)$ model);
here $\omega_n$ refers to Matsubara frequencies which are integral multiples of $2 \pi T$, but the conductivity is defined at all $\omega$ by
analytic continuation.
To make contact with the condensed matter literature, we have explicitly displayed a factor of the quantum unit of conductance 
\beq
\sigma_Q = \frac{(e^\ast)^2}{\hbar}\,, 
\eeq
where $e^\ast$ is the effective charge of the carriers 
($e^\ast = 2e$ for the superfluid-insulator transition of Cooper pairs); the ratio $\sigma/\sigma_Q$ is then a dimensionless
function whose values we will
present here. 
Note that, in the condensed matter literature $(e^\ast)^2/h=2\pi\sigma_Q$ is often used as a definition of the quantum unit of conductance.

The OPE specifies the behavior of the product of a pair of operators when they approach the same point in spacetime:
the product is replaced by a sum over the operators of the CFT with universal coefficients \cite{Rychkov,Rychkov2}. These OPE coefficients ultimately allow one to compute
all local correlators of the CFT at $T=0$. At $T>0$, the OPE expansion is applicable for times $|t| < \hbar/(k_B T)$ (we will set $\hbar = k_B =1$ in subsequent expressions), 
but cannot be used
directly for longer times which are naturally sensitive to the global topology of spacetime, and in particular to the periodic boundary
conditions along the Euclidean temporal direction.
For our purposes, it is useful to work in frequency space, and to 
express the OPE as the product of 2 operators when they carry a common large Euclidean frequency. 
One of our primary results is the following OPE of the product of 2 currents
\bea
\lim_{|\Omega| \gg p} J_x (\bomega) J_x (-\bomega+\bp) &=& - |\Omega| \, \sigma_\infty \, \delta^{(3)} (\bp) - \frac{\mathcal{C}}{|\Omega|^{\Delta-1}} 
\mathcal{O} (\bp) \nn
&~&~
+\frac{\mathcal{C}_{T}}{\Omega^2}
\Bigl[ T_{xx} (\bp) - T_{yy} (\bp) -12 \gamma (T_{xx} (\bp) + T_{yy} (\bp)) \Bigr] + \dotsb, \label{eq:mainope}
\eea 
where $\bomega \equiv (\Omega, 0, 0)$, with $\Omega$ being the imaginary frequency at $T=0$, and $\bp$ is a fixed 3-momentum with $p \equiv |\bp|$. The structure of this OPE was deduced by computing correlators of the operators on the left-hand-side
with those on the right-hand-side using the $1/N$ expansion of the O($N$) model; it is also consistent with correlators deduced from holography.
Taking an expectation of the above equation at any temperature
will lead to both sides being proportional to $\delta^{(3)}(\bp)$.
Here $\sigma_\infty$ is limiting value of the conductivity obtained as $T\ra 0$, $\mathcal{O}$ is a possible scalar operator in the OPE with
scaling dimension $\Delta$, $T_{\mu\nu}$ is the energy-momentum tensor, and
$\mathcal{C}$, $\mathcal{C}_T$, and $\gamma$ are OPE coefficients.

The terms in \req{mainope} involving the energy-momentum tensor have been implicitly included in previous 
studies \cite{myers11,suvrat,krempa_nature}. In the holographic approach, these terms arise from
the coupling, $\gamma$, of the Weyl tensor to the gauge flux \cite{ritz,myers11}; 
the value of $\gamma$ obeys the exact bound \cite{myers11,suvrat} 
$|\gamma| \leq 1/12$. It is also interesting to note the resemblance of the energy-momentum terms in \req{mainope} to the
Sugawara construction \cite{sugawara,Goddard85} of the energy-momentum tensor from the OPE of currents in CFTs in 1+1 dimension;
indeed, the term proportional to $\gamma$ is $T_{xx} + T_{yy} = - T_{\tau\tau}$, the Hamiltonian density. 

We can use \req{mainope} to determine the frequency dependence of the conductivity at finite temperature in the regime
$\hbar |\omega_n| \gg k_B T$, where $\omega_n$ is the Matsubara frequency 
(we will henceforth set $\hbar=k_B = 1$). We simply evaluate the expectation value
of the right-hand-side in an equilibrium thermal ensemble defined by the CFT, and indeed we have only displayed terms in
\req{mainope} which have a non-zero expectation value at $T>0$. By this method we obtain from 
Eqs.\ts(\ref{eq:condJJ}) and (\ref{eq:mainope})
\beq
\frac{\sigma (i\omega_n)}{\sigma_Q} = 
\sigma_{\infty} + b_1\, \left( \frac{T}{\omega_n} \right)^{\Delta} + b_2 \,\left( \frac{T}{\omega_n} \right)^3 
   + \dotsb, \quad \quad  \omega_n \gg T\,, \label{eq:sigmalargew}
\eeq
where the dimensionless numbers $b_1$, $b_2$ are related to the OPE coefficients $\mathcal{C}$ and $\mathcal{C}_T$ respectively.
This expression shows that the term associated
with the operator $\mathcal{O}$ is important when there is a scalar operator with
a scaling dimension $\Delta < 3$. For the O($N$) Wilson-Fisher CFT there is indeed such an operator: it is the ``thermal'' operator
$\mathcal{O}_g$,
whose introduction breaks no symmetry and drives the CFT into a non-critical state. We note that the label ``thermal'' 
descends from critical phenomena terminology, and is not meant to imply that $\mathcal{O}_g$ introduces a non-zero $T$;
such an operator has a coupling $g$ in the action, and $g$ has to be tuned to a critical value $g=g_c$ to realize the CFT.
The operator $\mathcal{O}_g$ has scaling dimension $\Delta$ which takes the value
\beq
\Delta_g = 3 - 1/\nu\,, \label{eq:deltanu}
\eeq
where $\nu$ is the correlation length exponent. For $N=2$, we have $\nu \approx 2/3$, and so the $\mathcal{O}=\mathcal{O}_g$ term
in Eqs.\ts(\ref{eq:mainope}) and (\ref{eq:sigmalargew}) is more important than that due to the energy-momentum tensor, at least at large $\omega_n$.

The previous analysis \cite{krempa_nature} did not allow for an operator $\mathcal{O}$ with $\Delta < 3$. 
Indeed, there is no such operator for numerous physically
interesting CFTs involving Dirac fermions coupled to gauge fields, including 
QED3. For these CFTs, the analysis of Ref.\ts\cite{krempa_nature} can be used
without modification.
However, for 
O($N$) Wilson-Fisher CFT, it is
necessary to extend the analysis to include the relevant operator $\mathcal{O}_g$; such an extension was briefly noted in Ref.\ts\onlinecite{ws}, but its
consequences were not appropriately analyzed.
After such an extension here, we find excellent compatibility between
Monte Carlo, operator product expansions, and holography, without any ad hoc rescaling of temperature.

We will begin our analysis by computations in the vector $1/N$ expansion for the O($N$) Wilson-Fisher CFT in Section~\ref{sec:on}.
With many details relegated to the Appendix, we obtain results for OPE coefficients and thermal expectation values.
Section~\ref{sec:qmc} presents our Monte Carlo results on the $N=2$ Wilson-Fisher CFT, and compares them with the $1/N$ expansion.
Section~\ref{sec:hol} turns to holography: by matching the large frequency behavior with the Monte Carlo results, we
are able to extrapolate to low frequency properties of the conductivity. Section~\ref{sec:fermions} presents a few results for CFTs
with Dirac fermions. Finally, in Section~\ref{sec:sr} we use the OPE analysis to prove conductivity sum rules.  

We close this introduction by summarizing our notations for the operators under consideration in Table~\ref{tab:op}, as they appear in 
Sections~\ref{sec:on}-\ref{sec:hol}.
\begin{table}[h]
  \centering
  \begin{tabular}{c||c|c|c|c} 
    $\mc O$ & $\De_{\mc O}$ & $\;\ell \;$ & $\ang{\mc O}_T$ & Holographic dual \\
    \hline
    $\phi_\al$ & $(1+\eta)/2$ & 0 & 0 & $-$ \\
    $\mc O_g \sim \phi_\alpha^2$ & $3-1/\nu$ & 0 & $B T^{3-1/\nu}$  & $\varphi$ \\
    $J_\mu$ & 2 & 1 & 0 & $A_\mu$ \\
    $T_{\mu\nu}$ & 3 & 2 & $H_{\mu\nu}T^3$ & $g_{\mu\nu}$
  \end{tabular}
  \caption{Main operators of the CFT describing the O$(N)$ Wilson-Fisher CFT in 2+1 dimensions. $\De_{\mc O},\ell$ 
are the scaling dimension and spin of the operators, respectively. 
The properties of the
conserved current $J_\mu$ (with flavor index suppressed) and energy-momentum tensor $T_{\mu\nu}$ are general for any CFT in 2+1 dimensions.} 
\label{tab:op}
\end{table} 

\section{O($N$) CFT}
\label{sec:on}

The theory of primary interest to us is described by the partition function for a O$(N)$ vector field $\phi_\alpha$, $\alpha = 1,\dotsc,N$,
\beq
Z = \int \mathcal{D} \phi_\alpha \exp \left( -
\int_\bx  \left[ \frac{1}{2} (\partial \phi_\alpha)^2  + \frac{v}{2N} \left( \phi_\alpha^2  - N/g\right)^2 
 \right] \right)\,, \label{eq:Z}
\eeq
where $\int_\bx \equiv \int d^3 x$ is the integral over 2+1 dimensional spacetime, 
$v$ parametrizes the quartic non-linearity, and $g$ is the tuning parameter across a quantum phase
transition between phases where O($N$) symmetry is broken and present. We have written this field theory in a somewhat unconventional
notation to facilitate a $1/N$ expansion; to the extent possible, we follow the notation in Ref.\ts\cite{podolskyss}.
In the limit $v \rightarrow \infty$ this theory reduces to the O($N$) non-linear sigma model. However, it is a subtle matter to identify
the thermal operator in the strict $v=\infty$ theory, as was discussed in Ref.\ts\cite{podolskyss}. We will therefore keep $v$ finite
for now, but will shortly indeed take the $v \rightarrow \infty$ limit when it no longer interferes with the scaling limit.

We will primarily be interested in the conductivity of this theory at the quantum critical point $g=g_c$ as a function of frequency, $\omega$,
and absolute temperature $T$. Without loss of generality, we focus on one of the conserved O($N$) currents of this theory, 
\beq
J_x = \phi_1 \partial_x \phi_2 - \phi_2 \partial_x \phi_1\,.
\eeq
The computationally challenging regime is at low frequencies $|\omega| \ll T$,
where we have the dissipative dynamics of the CFT relaxing to thermal equilibrium. However, controlled and reliable studies
are possible at high frequencies $\omega \gg T$. In this section, we will present the results of a $1/N$ expansion of the behavior
of the conductivity in this $\omega \gg T$ regime using the OPE in \req{mainope}.

The leading term in \req{mainope} is given by the constant $\sigma_\infty$ which has been computed earlier.
For completeness, we note its value in the $1/N$ expansion \cite{Cha91,yejin} for the theory $Z$
\beq
\sigma_\infty = \frac{1}{16} \left( 1 - \frac{1}{N} \frac{64}{9 \pi^2} + \mathcal{O}(1/N^2) \right).
\label{Cha}
\eeq

The terms in \req{mainope} involving the energy-momentum tensor have been discussed previously 
in different formulations \cite{myers11,suvrat,krempa_nature}.  We can use holography to compute the 3-point
correlator between $J_x$, $J_x$, and $T_{\mu\nu}$ as described in Ref.\ts\cite{suvrat}, and then deduce the structure of the
OPE: this computation in described in Appendix~\ref{app:t}. We can also compute the same 3-point correlator in the $1/N$ expansion
as described in Ref.\ts\cite{suvrat}, and again obtain \req{mainope} with specific values of the OPE coefficients: this is also described
in Appendix~\ref{app:t}. The $1/N$ expansion for $\gamma$ for the theory $Z$ is \cite{suvrat} 
\beq
\gamma = - \frac{1}{12} + \mathcal{O} (1/N)\,. \label{eq:gammaval}
\eeq
Similarly, the $1/N$ expansion for $\mathcal{C}_T$ is  
\beq
\mathcal{C}_T = \frac{4}{N} + \mathcal{O}(1/N^2)\,. \label{eq:CTval}
\eeq

For the O($N$) field theory in \req{Z}, there is a relevant scalar operator $\mathcal{O}$ which we denote $\mathcal{O}_g$ because 
it is generated by tuning $g$ away from the quantum critical point. This is the operator $\mathcal{O}_g \sim \phi_\alpha^2$
with scaling dimension in \req{deltanu}.

We will compute the OPE coefficient of $\mathcal{O}_g$ in the $1/N$ expansion of $Z$. An important subtlety arises in the definition
of $\mathcal{O}_g$ in such an expansion, as we now describe. The scaling limit of the large $N$ expansion also involves taking
the limit \cite{brezin} $v \rightarrow \infty$ in the action in \req{Z}. However, in this limit, we see from \req{Z} that $\phi_\alpha^2 = N/g$, a constant. Consequently, the correspondence $\mathcal{O}_g \sim \phi_\alpha^2$, assumed in Ref.\ts\onlinecite{ws}, does not define an appropriate non-constant thermal operator
at $v=\infty$.
A proper definition of $\mathcal{O}_g$ requires a more careful analysis of the $N\rightarrow \infty$ and $v \rightarrow \infty$ limits
\cite{podolskyss}. We decouple the quartic
term in $Z$ by a Hubbard-Stratonovich field $\widetilde{\lambda}$ and write
\begin{equation}
Z = \int \mathcal{D} \phi_\alpha \mathcal{D} \widetilde{\lambda} \exp \left( -
\frac{1}{2} \int_\bx  \left[ \left(\partial \phi_\alpha\right)^2
 + \frac{i}{\sqrt{N}} \widetilde{\lambda} \left(\phi_\alpha^2 - N/g\right) + \frac{\widetilde{\lambda}^2}{4 v}  \right] \right) \,.
 \label{eq:Z1}
\end{equation}
It is the field $i \widetilde{\lambda}$ which we will identify with the operator $\mathcal{O}_g$. 
This identification is motivated by the following identities between the one- and two-point correlators of $\phi_\alpha^2$ and $\widetilde{\lambda}$ (which
can be obtained by taking appropriate functional derivatives of source terms) \cite{podolskyss}
\bea
\left\langle \phi_\alpha^2 (\bx) \right\rangle &=&  \frac{N}{g} + i \frac{\sqrt{N}}{2v} \left\langle \widetilde{\lambda} (\bx) \right\rangle\,; \nn
\left\langle \phi_\alpha^2 (\bx) \phi_\beta^2 (0) \right\rangle - \left\langle \phi_\alpha^2 (0) \right\rangle^2 &=& \frac{N}{v} \delta^{(3)} (\bx) - \frac{N}{4v^2} \left[
\left\langle \widetilde{\lambda} (\bx) \widetilde{\lambda} (0) \right\rangle - \left\langle \widetilde{\lambda} (0) \right\rangle^2 \right]. \label{eq:amplam}
\eea 
So up to unimportant additive terms, the correlators of $\phi_\alpha^2$ are equal to those of $(\sqrt{N}/(2 v)) i \widetilde{\lambda}$.
As reviewed in Ref.\ts\cite{podolskyss}, the correlators of $i \widetilde{\lambda}$ have
a sensible scaling limit in a theory in which we take the $v \rightarrow \infty$ limit already in the action in 
\req{Z1}. So we identify $\mathcal{O}_g \sim i \widetilde{\lambda}$, and then set $v=\infty$ in subsequent computations.
Note, however, that Eqs.\ts(\ref{eq:amplam}) become trivial at $v = \infty$, and so $v$ has to be kept finite only in deducing the correlators
of $\phi_\alpha^2$.
Specifically, we define the ``thermal'' operator by
\beq
\mathcal{O}_g (\bx) = C_\lambda \, i \widetilde{\lambda} (\bx)\,, \label{CT}
\eeq
where the cutoff-dependent constant $C_\lambda$ will be chosen so that the two-point correlator of $\mathcal{O}_g$ is normalized as
\beq
\left\langle \mathcal{O}_g (\bp) \mathcal{O}_g (-\bp) \right\rangle - \left\langle \mathcal{O}_g \right\rangle^2 = - 16 p^{3-2/\nu}.
\label{eq:otot}
\eeq
The pre-factor of 16 is chosen for convenience in the $1/N$ expansion: we find in Appendix~\ref{app:t0} that $C_\lambda = 1$
at $N=\infty$. With these definitions and normalizations, we can compute the value of the OPE coefficient $\mathcal{C}_g$: we find
in Appendix~\ref{app:t0} that
\beq
\mathcal{C}_g = \frac{1}{4 \sqrt{N}} + \mathcal{O} (1/N^{3/2})\,. \label{eq:Cgval}
\eeq

With all the ingredients in the OPE at hand, we can proceed to the determination of $T>0$ behavior of the conductivity 
from \req{mainope}. For this, we need the expectation values of $\mathcal{O}_g$ and $T_{\mu\nu}$ at $T>0$:
these are determined in Appendix~\ref{app:gt0}. 

For the operator $\mathcal{O}_g$ we obtain
\beq
\left\langle \mathcal{O}_g \right\rangle_T - \left\langle \mathcal{O}_g \right\rangle_{T=0} \equiv B
T^{3 - 1/\nu}  
\label{eq:ogval2}
\eeq 
with
\beq
B = \sqrt{N} \Theta^2 \left[1  - \frac{1.8914}{N} + \mathcal{O}(1/N^2) \right],
\label{eq:Bval}
\eeq
where
\beq
 \Theta \equiv 2 \ln \left( \frac{\sqrt{5}+1}{2} \right).
 \label{eq:Thetaval}
\eeq 
In \req{ogval2}, we note that we have subtracted the $T=0$ expectation value, which is non-universal
and finite in the field theory \req{Z}. This subtraction can be seen as defining the scaling operator of the 
IR fixed point CFT, in which the expectation values at zero temperature vanish.
For the purposes of comparing with our Monte Carlo results, it useful to express this result in a form that
is independent of our arbitrary normalization of $\mathcal{O}_g$ in \req{otot}.
We take the Fourier transform of \req{otot} to real space to obtain  
\beq
\left\langle \mathcal{O}_g (\bx) \mathcal{O}_g (0) \right\rangle \equiv \frac{A}{x^{6-2/\nu}} \label{eq:defA} 
\eeq
with
\beq
A = - \frac{2^{7-2/\nu} \Gamma(3-1/\nu)}{\pi^{3/2} \Gamma(-3/2+ 1/\nu)}\,.
\label{eq:Aval} 
\eeq
From Eqs.\ts(\ref{eq:Aval}) and (\ref{eq:Bval}) we can construct the universal ratio which is independent of the 
normalization convention of $\mathcal{O}_g$:
\beq
\Upsilon = \frac{\sqrt{A}}{B} = \frac{4}{\pi \Theta^2 \sqrt{N}} \left[ 1 + \frac{0.8941}{N} + \mathcal{O}(1/N^2)  \right] . \label{eq:valUpsilon}
\eeq
This ratio will be compared with quantum Monte Carlo results for $N=2$ in Section~\ref{sec:qmc};
its value will also be useful in the holographic analysis in Section~\ref{sec:hol}.

For the $T>0$ expectation value of the energy-momentum tensor, we have for any CFT
\beq
\left\langle T_{xx} \right\rangle_{T} = \left\langle T_{yy} \right\rangle_{T} 
= - \frac{1}{2} \left\langle T_{\tau\tau} \right\rangle_{T}
= H_{xx} T^3 \,, 
\eeq
which corresponds to the pressure of the CFT.
We have implicitly subtracted from these expectation values their $T=0$ value; $H_{xx}$ is a universal number characterizing the CFT. This equation manifestly shows
the tracelessness of $T_{\mu\nu}$ in a CFT, which holds at finite temperature. 
The computation in Appendix~\ref{app:gt0} shows that in the large-$N$ limit of the O($N$) model
\beq 
H_{xx} 
= \frac{\zeta(3)}{2\pi}\left(\frac{4N}{5}-0.3344\right)  \, . 
\label{eq:Tval2}
\eeq
Collecting our results, we can now insert Eqs.\ts(\ref{eq:ogval2}) and (\ref{eq:Tval2}) into \req{mainope} and obtain the large frequency
behavior of the conductivity in the O($N$) CFT:  
\begin{align}
\frac{\sigma (i\omega_n)}{\sigma_Q} = \sigma_{\infty} + \mathcal{C}_g \, B \, \left( \frac{T}{\omega_n} \right)^{3-1/\nu} + 24 \, \mathcal{C}_T \, \gamma \, H_{xx} \,\left( \frac{T}{\omega_n} \right)^3 + \dotsb \,. \label{eq:sigmaT}
\end{align}
Note that this is the result for Euclidean frequencies $\omega_n \gg T$.
We show in Appendix~\ref{app:gt0} that the result agrees precisely with explicit
computation of the conductivity in the $N=\infty$ theory, which appears in \req{sigmaseries}.
The result Eq.~(\ref{eq:sigmaT}) shows that the combination $\mathcal{C}_g B$ is also independent of the normalization convention 
of $\mathcal{O}_g$.

The analytic continuation $i\w_n\ra\w+i0^+$ of \req{sigmaT} to real frequencies $\w \gg T$ yields:
\begin{align}
  \frac{\s(\w/T)}{\s_Q} = \s_\infty + b_1 \left[\re(i^{\De_g})+i\im(i^{\De_g})\right] \left( \frac{T}{\w} \right)^{\De_g} 
  - i\,b_2 \left( \frac{T}{\w} \right)^3 +\dotsb,
\end{align}
where $\De_g=3-1/\nu$. We note that for finite $N>1$, the scaling dimension $\De_g$ is not an integer, making both the
real and imaginary parts of $\s(\w/T)$ to scale like $(T/\w)^{\De_g}$ at large $\w/T$. For instance if we set $\nu=2/3$, 
this yields $\De_g=3/2$, thus $\im(i^{3/2})=-\re(i^{3/2})=1/\sqrt 2$. In contrast, the $N=\infty$ limit is special because 
$\De_g$ is an integer, 2, and thus
only the real part scales like $(T/\w)^2$, while the imaginary part decays faster, \ie as $(T/\w)^3$.
In the case of CFTs that do not have a scalar operator with scaling dimension $\De<3$ in the $JJ$ OPE, the real and imaginary parts
of the conductivity at asymptotically large and real frequencies behave differently. The imaginary parts decays
as $(T/\w)^3$ due to the energy-momentum tensor, while the real part decays faster due the presence of other operators.
This is the case for certain CFTs involving Dirac fermions discussed in Section~\ref{sec:fermions}.
It is also the case for the holographic models previously considered \cite{ritz,myers11}, as shown in Ref.\ts\cite{william2}.

\section{Quantum Monte Carlo}
\label{sec:qmc}

In order to perform efficient Quantum Monte Carlo simulations of \req{Z}  for $N=2$ it is useful to 
introduce a simple lattice model in the same  
universality class. For this purpose we use a quantum rotor model
defined in terms of phases $\theta_{\vec{r}}$ living on the sites,  $\vec{r}$, of a two-dimensional square lattice:
\begin{equation}
    H_{\text{qr}}=\frac{U}{2}\sum_{\vec{r}} 
    \frac{1}{2}\left( \frac{1}{i}\frac{\partial}{\partial
            \theta_{\vec{r}}} \right)^2-\mu\sum_{\vec{r}}\frac{1}{i}\frac{\partial}{\partial
              \theta_{\vec{r}}}
              -\sum_{\langle \vec{r},\vec{r}'\rangle }
              t \cos(\theta_{\vec{r}}-\theta_{\vec{r}'}) \ .
              \label{eq:hqr}
\end{equation}
Here $-i{\partial}/{\partial \theta_{\vec{r}}}$ is usually identified with the angular momentum of the quantum rotor at site ${\vec r}$, which is the canonical conjugate
of $\theta_{\vec{r}}$. However,  it can also be viewed as the deviation from an average (integer) particle
number and this model is therefore in the same universality class as the Bose Hubbard model. 
The on-site repulsive interaction, $U$, then hinders large deviations from the mean particle number while $t$ characterizes the hopping between nearest neighbor sites. 
For completeness, we include a chemical potential $\mu$ although the case of integer filling that we focus on here corresponds to $\mu=0$. 

As discussed in Ref.\ts\onlinecite{krempa_nature}, it is possible to directly simulate Eq.~(\ref{eq:hqr}) using quantum Monte Carlo (QMC) techniques. However, it is useful
to further simplify the model by employing the Villain approximation~\cite{Villain} where the $\cos\theta$ term is replaced by  a sum of
periodic Gaussians centered at $2\pi m$ (where $m$ is an integer):
$
\exp({t\Delta\tau\cos(\theta)})\simeq \exp({t\Delta\tau})\sum_m\exp({-\frac{1}{2}t\Delta\tau
    (\theta - 2\pi m)^2}),
$
preserving the periodicity of the Hamiltonian in $\theta$.
A standard Trotter decomposition can then be performed where $\beta U$, is divided into $L_\tau$ slices of size $\Delta\tau=\beta U/L_\tau$. One then
arrives at a model defined in terms of  an {\it integer-valued} current ${\bf J}=(J^\tau,J^x,J^y)$ with $J^\tau$ the angular momentum (or particle number) living on the links
of a $2+1$ dimensional discrete lattice of dimensions $L\times L\times L_\tau$:~\cite{Cha91,sorensen92,Wallin94} 
\begin{equation}
  Z_{V}\approx 
{\sum_{\{\bf J\}}}'
\exp\left[-\frac{1}{K}
\sum_{(\tau,{\vec r})}\left( 
    \frac{1}{2}{\bf J}^2_{(\tau,{\vec r})}-\frac{\mu}{U} J^\tau_{(\tau,{\vec r})}   
    \right)\right] \ . 
\label{eq:ZV}
\end{equation}
Here $L_\tau\Delta\tau$ takes the place of the dimensionless inverse temperature $\beta U$ and varying $K$ is analogous to varying $\sqrt{t/U}$ in the
quantum rotor model. We stress that, the ${\sum}'$ denotes the fact that the summation over ${\bf J}$ is constrained to divergence-less configurations making the
summation over the integer valued currents highly non-trivial to perform. In deriving the Villain model a fixed $\Delta\tau=1/K$ is used. Despite the fixed, rather large, value
of $\Delta\tau$, the Villain model has several significant advantages. Most notably, it is explicitly isotropic in space
and time. Secondly, very efficient Monte Carlo algorithms have been developed for the Villain model~\cite{aleta,aletb} as well as for the quantum rotor model. Here we
use {\it directed} Monte Carlo techniques as described in Ref.\ts\onlinecite{aletb}. 
The location of the QCP is also known, $K_c=0.3330671(5)$.~\cite{pollet,krempa_nature} Further details of the numerical calculations are given in Appendix~\ref{app:qmc}.

In the condensed matter literature the quantum of conductance is usually defined as $(e^*)^2/h$ (for carriers of charge $e^*$), however, here we use a slightly different definition of
$\sigma_Q=(e^*)^2/\hbar$ that is also widely used. 
In terms of $\sigma_Q$, the frequency dependent 
conductivity of the Villain model can then be calculated by evaluating ($\omega_n$ are the Matsubara Euclidean frequencies)
\begin{equation}
\frac{\sigma(i\omega_n)}{\sigma_Q} = 
\frac{1}{L^{d-2} 2\pi n}\left\langle \left|\frac{1}{L}\sum_{(\tau,{\vec r})}e^{i\omega_n \tau} J^x_{(\tau,{\vec r})}\right|^2\right\rangle\,, 
  \end{equation}
  which is dimensionless in $d=2$. Here $n$ is an integer labeling the (dimensionless) Matsubara frequency $\omega_n/\omega_c=2\pi n/L_\tau$. We note that this
  expression is explicitly independent of the imaginary time discretization $\Delta\tau$ even though $\omega_n$ is measured in units of $\omega_c=U/\Delta\tau$ 
  and any residual dependence of $\sigma(i\omega_n)$ on $\Delta\tau$ is therefore
  usually ignored.

\begin{figure}   
\centering
\includegraphics[scale=.65]{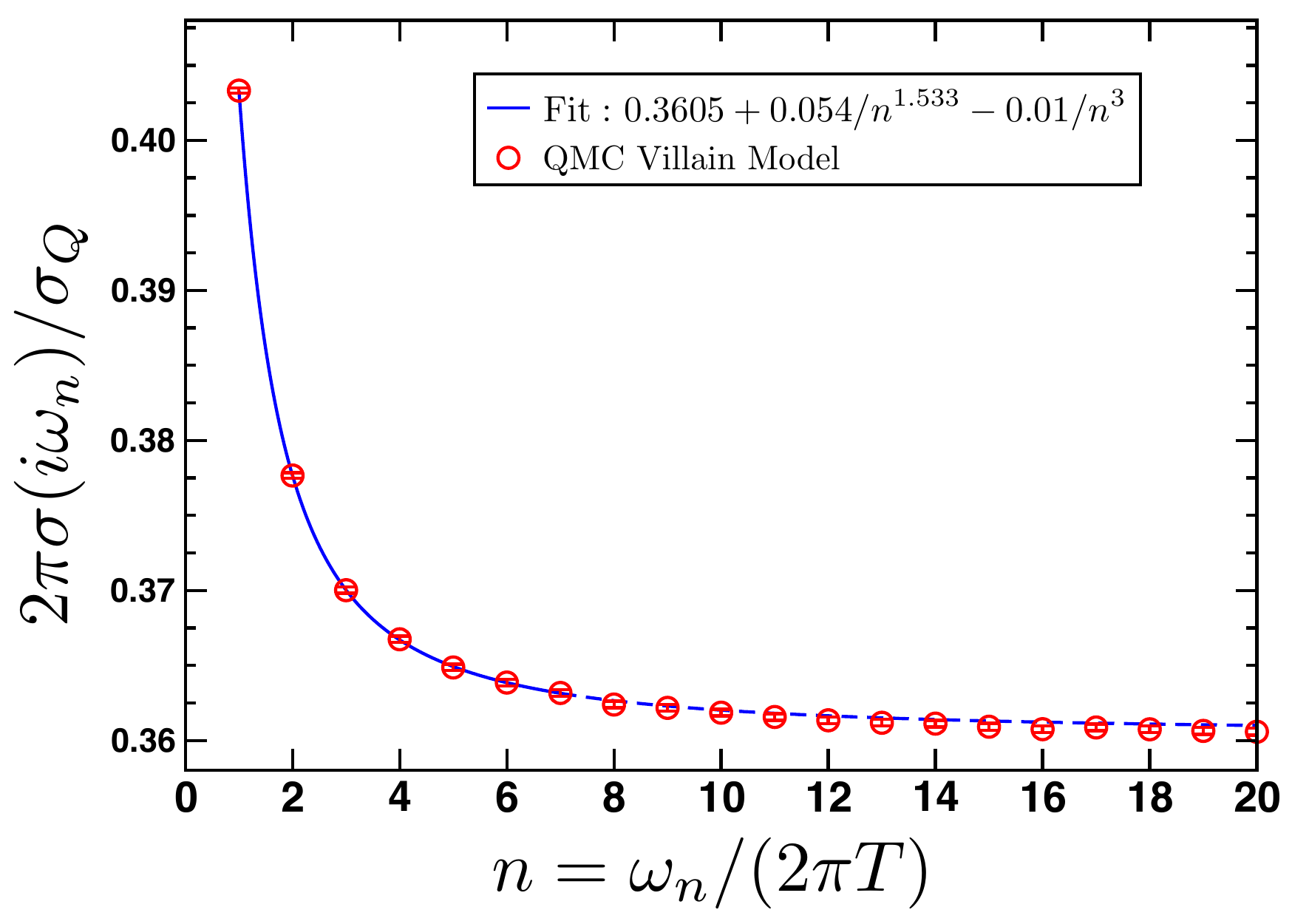}   
\caption{\label{fig:Sigma_V_large_n} QMC results (open circles) 
  at $K_c=0.3330671$ with $\mu=0$ for the frequency dependent conductivity $\sigma(i\omega_n)$. All
    results have first been extrapolated to $L\to\infty$ and subsequently to $T\to 0$ ($L_\tau\to\infty$).
    The solid blue line shows a fit to the QMC data for $n=1,\ldots,7$ of the form $2\pi\sigma/\sigma_Q=0.3605+0.054/n^{1.533}-0.01/n^3$ with $n=\omega_n/(2\pi T)$ the Matsubara index. 
    The dashed blue line is the continuation of the fitted form to $n>7$.   
} 
\end{figure}
The conductivity at the QCP has previously been studied~\cite{Cha91,sorensen92,Wallin94}. The first attempts at calculating the universal $T\to 0$ limit
of the conductivity~\cite{sorensen} appeared significantly later and the first large scale numerical calculations of this quantity 
have only very recently been performed~\cite{krempa_nature,pollet} due to their extremely demanding nature. Here we re-analyze the numerical results of 
Ref.\ts\onlinecite{krempa_nature}
in order to test the analytical result, Eqs.\ts(\ref{eq:sigmalargew}) and (\ref{eq:sigmaT}). The $T\to 0$ extrapolated QMC results for the conductivity are shown
in Fig.\ts\ref{fig:Sigma_V_large_n} along with our fit. 
For a discussion of the numerical details of the $T\to 0$  extrapolation we refer to the supplementary material of Ref.\ts\onlinecite{krempa_nature} as well as to
Appendix~\ref{app:qmc}.
Performing the $T\to 0$ extrapolation for large values of $n$ is significantly more difficult than at small values of $n$. We have therefore limited the values of $n$ that
we use in the fit to $n=1,\ldots,7$ where we have the highest confidence in the $T\to 0$ extrapolated QMC results. For these values of $n$ we obtain remarkably good agreement between
the fit and the QMC results. Furthermore, as can be seen, the fit works very well also
for $n>7$.
We note that, even though values of $n=1\ldots 7$ used in the fit in \rfig{Sigma_V_large_n} may appear rather small, they correspond to values of $\omega_n/T\geq 2\pi$ 
where Eqs.\ts(\ref{eq:sigmalargew}) and (\ref{eq:sigmaT}) should be applicable.
Inserting appropriate powers of $2\pi$, the fit in \rfig{Sigma_V_large_n} can be converted to a fit to \req{sigmalargew}
and we find fitted values of $\sigma_\infty$, $\nu$, $b_1$, and $b_2$ as follows 
\bea
2 \pi \sigma_\infty &=& 0.3605(3)  \nn
\nu &=& 0.68(3) \nn
b_1 &=& 0.143 (5) \nn
b_2 &=& -0.4(1)\,,
\label{fits}
\eea
where we only quote statistical errors arising from the fit. 
We comment on these values in turn: 
\begin{itemize}
\item
The value of $2 \pi \sigma_\infty$ is in excellent agreement with existing results \cite{krempa_nature,pollet,gazit14}. 
Comparing with the large $N$ result in Eq.~(\ref{Cha}), the $N=\infty$ value is $0.39$, while the $1/N$ corrected expression evaluated
at $N=2$ yields $0.25$.
\item 
Our fit in \rfig{Sigma_V_large_n} provides a value for $\nu$ that is consistent with the much more precise estimate obtained in \rfig{OTV} (see below) as
well as with previous numerical studies \cite{Campostrini01,Burovski06,Campostrini06}.
\item For $b_1$, we can only compare with the $N=\infty$ result obtained in Section~\ref{sec:on}. From Eqs.\ts(\ref{eq:sigmaT}), (\ref{eq:Cgval}) 
and (\ref{eq:Bval}),
or equivalently from \req{sigmaseries},  
we obtain $b_1 = \Theta^2/4 = 0.23$.
\item Our fits to $b_2$, the coefficient of the $(T/\omega_n)^3$ term, are not accurate. But the presence of a {\it negative} $b_2$ can be 
reliably confirmed. Comparing with the $N=\infty$ results of Section~\ref{sec:on}, from Eqs.\ts(\ref{eq:sigmaT}), (\ref{eq:gammaval}), (\ref{eq:CTval}) and (\ref{eq:Tval2}),
or equivalently from \req{sigmaseries}, 
we obtain $b_2\big|_{N=\infty} = -1.2$. Using the $1/N$ correction for the pressure coefficient $H_{xx}$, \req{Tval2}, 
we get $b_2\approx -0.97$.  
Both the $1/N$ expansion and QMC simulations suggest a \emph{negative} $\gamma$ for the O(2)
CFT, which differs from the positive value extracted via the ``holographic continuation'' analysis done in Ref.\ts\onlinecite{krempa_nature}. The new holographic analysis performed in this work is consistent with a negative value of $\g$, because
it incorporates the relevant scalar operator $\mc O_g$. 
\end{itemize}

\begin{figure}   
\centering
\includegraphics[scale=.45]{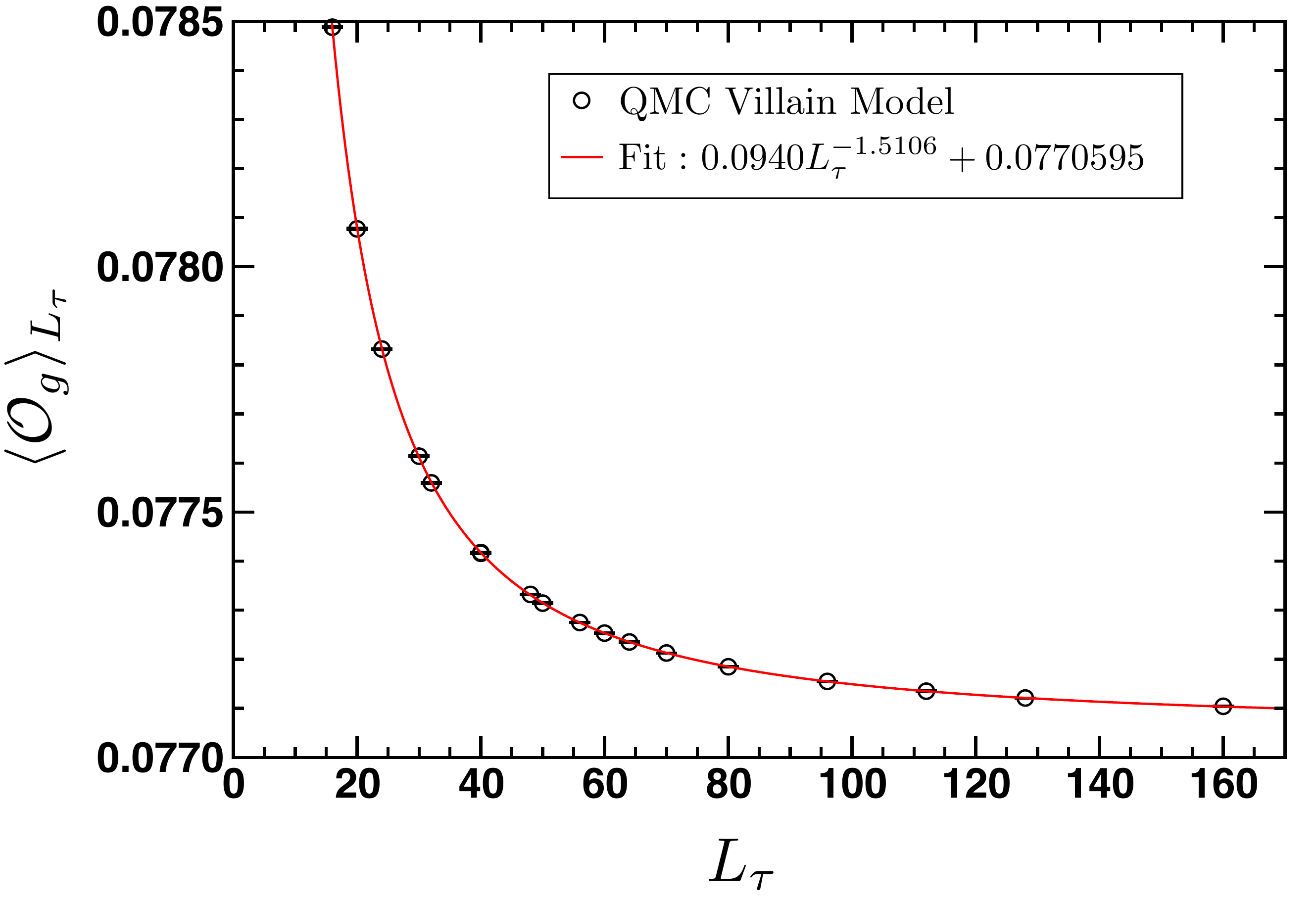} 
\caption{\label{fig:OTV} QMC results (open circles) at $K_c=0.3330671$ for $\langle \mathcal{O}_g \rangle$ in the limit $L\to\infty$ as a function of 
$L_\tau$. 
The solid red line indicates a fit to the QMC data of the indicated form. 
} 
\end{figure}   
Next we turn to correlations of the ``thermal'' operator $\mathcal{O}_g$. For the Villain model, it is convenient to define this operator by 
\begin{equation}
\mathcal{O}_g(\tau,{\vec r})=\frac{1}{2}{\bf J}^2_{(\tau,{\vec r})}-\frac{\mu}{U} J^\tau_{(\tau,{\vec r})}\,. \label{eq:villainog}
\end{equation}
By suppressing winding number fluctuations in the spatial directions and using system sizes with spatial dimensions $L>L_\tau$~\cite{pollet} it is possible
to effectively calculate $\langle \mathcal{O}_g \rangle$ in the limit $L\to\infty$ with finite $L_\tau$. Our results are shown in Fig.~\ref{fig:OTV}.
An extraordinary good agreement with the analytical expression \req{ogval2} is evident. The fit shown in Fig.~\ref{fig:OTV} immediately yields
\begin{equation}
\nu = 0.6714(10)\,,
  \label{eq:nu}
\end{equation}
in excellent agreement with other recent estimates~\cite{Campostrini01,Burovski06,Campostrini06} confirming that $\nu$ is slightly larger than $2/3$. In fact, the precision at which $\nu$ can be determined from $\langle \mathcal{O}_g \rangle$
makes this a promising venue for a future high precision determination of $\nu$.
Furthermore, from \rfig{OTV} we find that the coefficient $B$ in \req{ogval2} is
\begin{equation}
B=0.0940(6).
\end{equation}
Recall that the value of $B$ by itself is non-universal, and depends upon the microscopic choices we made in the definition in \req{villainog};
however we will combine it below with another observable to obtain a normalization-independent number.
For further analysis, it is also useful to note the non-universal value:
\begin{equation}
\langle \mathcal{O}_g \rangle_{L_\tau\to\infty}=0.0770595(5).
\label{eq:Oginf}
\end{equation}

\begin{figure}   
\centering
\includegraphics[scale=.5]{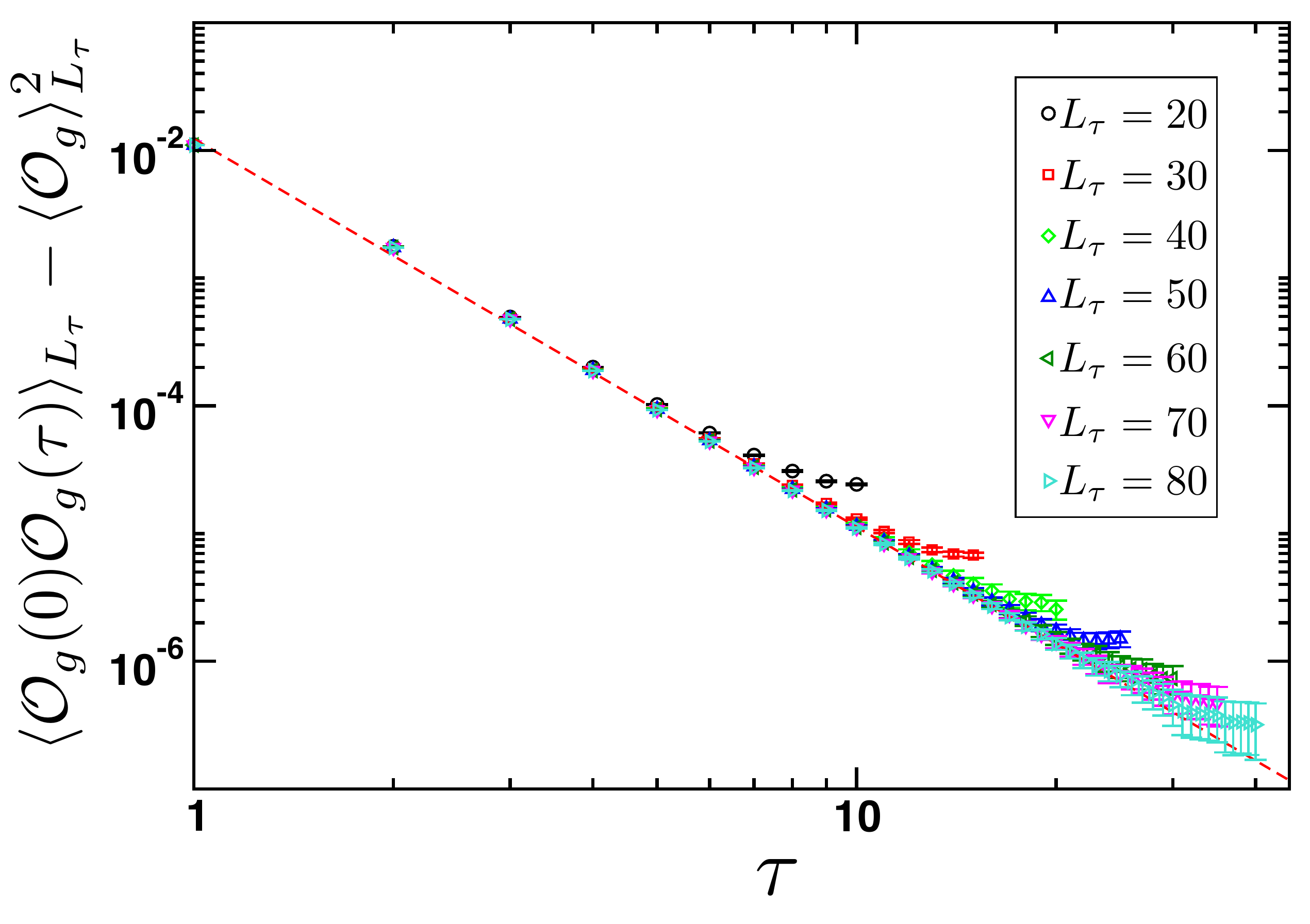}  
\caption{\label{fig:OTOTV} QMC results (open circles) at $K_c=0.3330671$ for $\langle \mathcal{O}_g(0) \mathcal{O}_g(\tau) \rangle_{L_\tau}-\langle \mathcal{O}_g\rangle^2_{L_\tau}$ 
  in the limit $L\to\infty$ as a function of the imaginary time, $\tau$. Results are shown for different values of $L_\tau$.
  The dashed red line indicates the $L_\tau\to\infty$ limit of $\langle \mathcal{O}_g(0) \mathcal{O}_g(\tau) \rangle_{L_\tau}-\langle \mathcal{O}_g\rangle^2_{L_\tau} \ \to\  0.0122 \tau^{-(6-2/\nu)}$
  with $\nu=0.6714$. 
} 
\end{figure}   
Next, we turn to the two-point correlation function of $\mathcal{O}_g$. Due to the space-time isotropy of the Villain model, 
it has the same behavior along the spatial 
and temporal directions. However, for convenience we focus on the temporal correlations. As before we perform calculations effectively in the $L\to\infty$ limit with a finite $L_\tau$.
Our results are shown in \rfig{OTOTV}. The data for individual values of $L_\tau$ are first fit to the form $A\left[\tau^{-(6-2/\nu)}+(L_\tau-\tau)^{-(6-2/\nu)}\right]+\langle \mathcal{O}_g\rangle^2_{L_\tau}$
    for $\tau>6$.
This yields values of $A$ that are close to independent of $L_\tau$ and we estimate:
\begin{equation}
A=0.0122(15)\,.
\end{equation}
The variations in $\nu$ in the fits are small, $\nu=0.671-0.675$, and consistent with the value of $\nu$ obtained above, \req{nu}. Furthermore, the fitted values for
$\langle \mathcal{O}_g\rangle^2_{L_\tau}$ are consistent with the actual calculated values of $\langle \mathcal{O}_g\rangle_{L_\tau}$ and clearly approach $\langle \mathcal{O}_g \rangle_{L_\tau\to\infty}^2$
as determined from \req{Oginf}. 

Finally, we can combine our computations of the one-point and two-point correlators of $\mathcal{O}_g$ to obtain a universal
number which is independent of the precise definition of $\mathcal{O}_g$ and the microscopic details of the action.
This is the ratio $\Upsilon$ defined in \req{valUpsilon}, and the present Monte Carlo studies yield: 
\begin{equation}
\Upsilon = \frac{\sqrt{A}}{B} = 1.18(13).
\label{eq:upqmc}
\end{equation}
Almost all of the uncertainty in this result arises from the uncertainty in the determination of $A$ which is difficult to calculate with high precision.
This result for $\Upsilon$ is in reasonable agreement 
with the $1/N$ expansion results in \req{valUpsilon}, where we have the $N=\infty$ value $\Upsilon = 0.97$, and the $1/N$ corrected
value at $N=2$ of $\Upsilon = 1.41$. 

We have also performed simulations directly of \req{hqr} which does not involve the Villain approximation. In this case it is considerably harder to obtain
high precision numerical data, however, our preliminary results indicate a value of $\Upsilon$ in very good agreement with the above results for the Villain model.

\section{Holography}
\label{sec:hol}

We have so far obtained systematic results for the conductivity in the high frequency regime $|\omega| \gg T$.
We also obtained quantum Monte Carlo results at the discrete Matsubara frequencies $\omega_n = 2 n \pi T$,
where $n$ is a non-zero integer. As we noted in Section~\ref{sec:intro}, we will now turn to holography to perform
the analytic continuation to all Minkowski frequencies.  

For the contributions of the energy-momentum tensor terms in \req{mainope}, such an analysis has already been
carried out in Ref.\ts\cite{krempa_nature}. So we turn to the extension needed to include the contribution 
of a scalar operator $\mathcal{O}$.

For the present purposes, the operator $\mathcal{O}$ is any operator in the OPE which obeys the analogs of the
Eqs.\ts(\ref{eq:ogval2}) and (\ref{eq:defA}) 
\bea
\left\langle \mathcal{O} (\bx) \mathcal{O} (0) \right\rangle &=& \frac{A}{x^{2 \Delta}} \quad, \quad T=0; \nn
\left\langle \mathcal{O} \right\rangle_T - \left\langle \mathcal{O} \right\rangle_{T=0} &=& B T^\Delta, \label{eq:Oratio}
\eea  
which define the normalization independent universal ratio $\Upsilon \equiv \sqrt{A}/B$. 

Now take the holographic dual of the same CFT in AdS$_{D+1}$ and the corresponding boundary operator
$\mathcal{O}(\bx)$ is represented by a bulk scalar field $\varphi(\bx, \tilde{u})$; here $\tilde{u}$ represents the emergent direction, and the AdS$_{D+1}$ metric
is $L^2 (d \bx^2 + d\tilde{u}^2)/\tilde{u}^2$ ($L$ is the AdS radius).
In the conventional normalization for the bulk scalar, the two-point correlator of $\mathcal{O}$ is \cite{suvrat}
\beq
\left\langle \mathcal{O} (\bk ) \mathcal{O} (-\bk) \right\rangle = - (2 \Delta - D) \frac{\Gamma (1- \Delta+D/2)}{\Gamma(1+ \Delta - D/2)} \left( \frac{k}{2} \right)^{2 \Delta - D}
\eeq
This translates in real space to
\beq
\left\langle \mathcal{O} ( \bx ) \mathcal{O} (0) \right\rangle = \frac{\pi^{-D/2} (D - 2 \Delta) \Gamma(\Delta) \Gamma(1-\Delta + D/2)}{\Gamma(D/2-\Delta) \Gamma(1+ \Delta - D/2)} \, \frac{1}{x^{2 \Delta}}
\eeq
For holography to reproduce the $T>0$ expectation values of the CFT with the same universal constant $\Upsilon$, we conclude from \req{Oratio} that 
\beq 
\left\langle \mathcal{O} \right\rangle_T - \left\langle \mathcal{O} \right\rangle_{T=0} = 
\frac{1}{\Upsilon} \left[ \frac{\pi^{-D/2} (D - 2 \Delta) \Gamma(\Delta) \Gamma(1-\Delta + D/2)}{\Gamma(D/2-\Delta) \Gamma(1+ \Delta - D/2)} \right]^{1/2}
 T^\Delta .
\eeq
Again using the standard AdS/CFT dictionary, we conclude that the bulk scalar must behave as 
(note that the metric is not modified at $T>0$ near the boundary $\tilde{u} \rightarrow 0$):
\bea
\varphi (\bx, \tilde{u}\rightarrow 0) &=& \frac{\tilde{u}^\Delta}{(2 \Delta-D)} \left( \left\langle\mathcal{O} \right\rangle_T - \left\langle\mathcal{O} \right\rangle_{T=0}\right) \nn
&=&
\frac{1}{\Upsilon (2 \Delta - D)} \left[ \frac{\pi^{-D/2} (D - 2 \Delta) \Gamma(\Delta) \Gamma(1-\Delta + D/2)}{\Gamma(D/2-\Delta) \Gamma(1+ \Delta - D/2)} \right]^{1/2}
(\tilde{u} \, T)^\Delta .\label{eq:e1} 
\eea
The $N=2$ Wilson-Fisher theory has $\Delta=\Delta_g$ given by \req{deltanu}
with $\nu = 0.67155(27)$ \cite{Campostrini01};
so 
$2 \Delta - D = 3 - 2/\nu \approx 0.02$ is nearly zero.
Fortunately, the coefficient in \req{e1} has a finite limit ($\approx 0.28 /\Upsilon$) as $\Delta \rightarrow D/2$.  

We now turn to deducing the consequences of the condensate of $\varphi$ in Eq.~(\ref{eq:e1}) at $T>0$.
Following the notation of Ref.\ts\onlinecite{ws}, it is convenient to introduce the dimensionless co-ordinate $u$, and the length scale $r_0$ by
\beq
u = \frac{\tilde{u} \, r_0}{L^2} \quad , \quad r_0 = \frac{4 \pi T L^2}{3}.
\eeq
Then the $T>0$ AdS$_4$-Schwarzschild metric is 
\begin{align}
  ds_{\rm Sch}^2=\frac{r_0^2}{L^2u^2}\left[-f(u)dt^2+dx^2+dy^2 \right]+\frac{L^2 du^2}{u^2 f(u)}\,,
\end{align} 
where 
\beq
f(u)=1-u^3 \, . \label{deff}
\eeq 
This spacetime is asymptotically ($u\ra 0$) AdS$_4$, with negative cosmological  
constant $\propto -1/L^2$, and contains a planar black hole with horizon at $u=1$.
We simplify notation for the near-boundary behavior of the field $\varphi$ in Eq.~(\ref{eq:e1}) by defining
\begin{align}\label{eq:phi-profile}
  \varphi(u\ra 0) = a u^\De+\cdots \,,
\end{align}
where the dots represent terms that decay faster as $u\ra0$, and $a$ is determined by the definitions above. 
The field
$\varphi$ will couple to the bulk gauge boson, $A_\mu$, dual to the current of the CFT like a dilaton, leading to the gauge action   
\beq 
\mc S=\int d^4x \sqrt{-g_{\rm Sch}} \left\{ -\frac{1}{4 g_4^2} [1 + \al\varphi (u) ] F_{ab}F^{ab} \right\}
\,, \label{eq:hol-gauge-lag}
\eeq
where $F_{ab}=\pd_a A_b-\pd_b A_a$, $g_4$ is the bulk gauge charge, and the coupling $\alpha$ is proportional to the OPE coefficient $\mathcal{C}$ in 
\req{mainope}.
As we shall see, the $\w_n\gg T$ asymptotic behavior of the conductivity of the corresponding boundary CFT is
\beq
\label{eq:sig-cft}
\frac{\sigma (i\omega_n)}{\sigma_Q} = \sigma_\infty + b_1 \left( \frac{T}{\omega_n} \right)^\Delta+\dotsb,
\eeq
where the dots denote subleading terms. The coefficient $b_1$ is as defined in \req{sigmalargew}, and it is proportional to 
the coupling $\alpha$ in \req{hol-gauge-lag}. As inputs to the holographic computation we will not use
the values of $\alpha$ and $\mathcal{C}$, but directly fit the value of $b_1$ to the Monte Carlo results in Eq.~(\ref{fits}).

Let us now determine the relation between $b_1$ and $a,\alpha$. In the $A_u=0$ gauge, the equation of motion which 
follows from \req{hol-gauge-lag} for the transverse component of the gauge field, $A_y$, 
(choosing $\vec k$ along the $x$-direction) is 
\beq  \label{eq:A-EOM}
\left((1+ \al\varphi)f A_y'\right)' - \mathfrak w^2 \frac{(1+\al\varphi)}{f} A_y =  0\, \quad; \quad
\mathfrak w \equiv \frac{3\w_n}{4\pi T}\,,  
\eeq
where we have defined $(\,)'=\pd_u(\,)$, and the rescaled imaginary frequency $\mathfrak w$. We note that
$\mathfrak w$ is defined for any value, not only at the discrete Matsubara frequencies.    
The function $f(u)$ appears in the metric, and was defined in Eq.~(\ref{deff}) 
(the results in this section hold for all $f(u)=1-u^p+\cdots$, with $p\geq 1$, so that the boundary metric is AdS$_4$).
To determine the power law $1/\w_n^\De$ in \req{sig-cft}, we can easily make use of the analysis of Ref.\ts\cite{william2},
which relies on the contraction map method employed in Ref.\ts\cite{sum-rules}. 
Here, we wish in addition to determine the coefficient $b_1$. 
This can be done perturbatively in $\alpha$, as we now show. It will be advantageous to
change the holographic coordinate from $u$ to $z$: $dz/du=1/f(u)$, \ie $z(u)=\int_0^u d\bar u/(1-\bar u^3)$. Note that
for $u\approx 0$, $z$ reduces to $u$.
Given the standard 
AdS/CFT prescription, the solution to \req{A-EOM} can be parameterized as $A_y = e^{-\mathfrak w z} + \alpha \tilde{A}$, with 
$\tilde{A}$ satisfying a Dirichlet condition at $z=0 (=u)$. To leading order in $\alpha$, $\tilde{A}$ obeys
\beq
\pd_z^2\tilde{A} -\mathfrak w^2 \tilde{A} =  \mathfrak w e^{-\mathfrak w z}\pd_z\varphi\,. 
\eeq
This equation can be solved by using a Green's function,
\beq
G(z,\bar z) = -\frac{1}{\mathfrak w} \left(\sinh(\mathfrak w z) e^{-\mathfrak w \bar z} ~\theta(\bar z-z) + \bar z \leftrightarrow z \right),
\eeq
where $\partial_z^2 G - \mathfrak w^2 G = \delta(z-\bar z)$. 
The current-current correlation function is then given by 
\beq
\langle J_x(\mathfrak w) J_x(-\mathfrak w) \rangle_{T} = \frac{1}{g_4^2} \pd_u A_y(u=0) = -\frac{\mathfrak w}{g_4^2}
\left(1 + \alpha \int_0^\infty dz~ e^{-2 \mathfrak w z} \pd_z\varphi +\dotsb \right).
\eeq 
Using the asymptotic behavior for the scalar profile, \req{phi-profile}, we obtain:  
\begin{align}
  \frac{\s(i\mathfrak w)}{\s_Q \s_\infty} &= -\frac{1}{\mathfrak w}\pd_uA(u=0) \\ 
  &= 1 + \al a \frac{\G(\De+1)}{2^\De} \frac{1}{\mathfrak w^\De}+\dotsb, \qquad\text{ for } \mathfrak w\gg 1\,.
\end{align}  
Comparing to \req{sig-cft} we find that we can indeed match the finite temperature CFT results, as long as 
\beq
\label{eq:b1}
b_1 = \sigma_\infty ~\al\, a \,\frac{\Gamma(\Delta+1)}{2^\Delta} \left(\frac{4\pi}{3} \right)^\De \,. 
\eeq
As a check, we can compare this result with the WKB analysis \cite{william2} done for the asymptotic behavior 
of $\s$ with a holographic model containing the term $\g L^2 C_{abcd}F^{ab}F^{cd}$. 
For the AdS$_4$-Schwarzschild metric, this term is also of the form given by Eqs.\ts(\ref{eq:phi-profile})
and (\ref{eq:hol-gauge-lag}), with $\alpha \, a = 4 \gamma$ and $\De=3$, which is the scaling dimension of the energy-momentum tensor. Then
the result above agrees
with the WKB analysis \cite{william2}: $b_1/\s_\infty=3\g\times (4\pi/3)^3$.  

We are now ready to use this relation in conjunction with simplest finite-temperature holographic model to determine the charge diffusion constant
and the conductivity at zero frequency. Here, it must be kept in mind that we are not including the long-time tails which were discussed in 
earlier work \cite{krempa_nature}. The full frequency dependence of the conductivity is discussed in Section \ref{sec:comparing}.  
 
\subsection{Holographic model for charge diffusion and conductivity} 

We shall proceed by examining the simplest holographic ansatz which models a CFT at finite temperature while reproducing its 
UV behavior. For this we simply assume that the $u \rightarrow 0$ behavior of the scalar profile in 
\req{phi-profile} holds all the way up to the horizon at $u=1$.
Such an ansatz connects naturally to the previous holographic analyses \cite{myers11,ritz,ws} that considered a four-derivative
term coupling the Weyl tensor to two field strengths, $\g L^2 C_{abcd}F^{ab}F^{cd}$:
for the AdS$_4$-Schwarzschild metric, this term has a $u^3$ dependence for all $u$, both near the
boundary $u \rightarrow 0$, and near the horizon $u \rightarrow 1$. We mention that in principle a more detailed
holographic analysis can be performed, where one determines the dilaton profile $\varphi(u)$ self-consistently with
the metric. It would be interesting to study the resulting IR behavior. We leave this for future investigation,
and proceed with our physically motivated ansatz, which, as we shall see, captures many essential features.

The charge diffusion constant for the background in \req{hol-gauge-lag} is well known and is given, for example in 
Refs.\ts\onlinecite{star1,myers11}: 
\beq
D = \frac{3}{4\pi T}[1+ \al\varphi(1)] \int_0^1 du ~\frac{1}{1+\al\varphi(u)}\,.
\eeq
Working perturbatively in $\alpha$ the above equation for the diffusion constant becomes  
\begin{align}
  D \approx \frac{3}{4\pi T}\left[1+ \alpha\,a ~\frac{\Delta}{\Delta+1} \right]= \frac{3}{4\pi T}\left[1+ \frac{b_1}{\sigma_\infty} ~\frac{\Delta}{\Gamma(\Delta+2)} \left(\frac{3}{2\pi}\right)^\Delta \right].
\end{align} 
From the last equality, we note that the growth of $b_1$ with $\De$ must be very rapid in order for an 
operator with large scaling dimension to make an important 
contribution to the charge diffusion constant, otherwise that operator will decouple. A similar statement can be made about the
d.c.\@\xspace conductivity:
\begin{align}
  \frac{\s(0)}{\s_Q} &= \frac{1}{g_4^2}[1+\al\phi(1)]= \frac{1}{g_4^2}(1+\al a)\,, \\
  &= \s_\infty + \frac{b_1 }{\G(\De+1)} \left(\frac{3}{2\pi}\right)^\De\,. \label{eq:sig-dc_large-Delta}
\end{align}
Before discussing the relevance of this analysis to generic CFTs, we point out an important caveat.  
Namely, that for generic CFTs we expect the conductivity to diverge logarithmically in the small
frequency limit $\w/T\ra 0$ due to long-time tails.  
This classical effect leads to the slow decay of correlators of conserved currents at long times; 
see the discussion in Refs.\ts\cite{ws,krempa_nature} for further details.   
Such long-time tails do not occur in the tree-level (or classical) holographic 
models that we consider due to an implicit limit of infinite number of CFT fields. Our 
holographic analysis therefore cannot describe the conductivity of the O(2) CFT when $\omega\ll T$.
(We point out that holography can capture long-time tails if $1/N$ quantum corrections are taken
into account \cite{saremi}.)
To circumvent the need to refer to long-time tails, one could replace the statements about $\w=0$,
such as \req{sig-dc_large-Delta},  
by equivalent statements at small but finite frequencies, say on the order of the temperature. 
The analysis above becomes more involved but we 
expect similar conclusions for the holographic model under consideration.

In a typical CFT once temperature is turned on there will be an infinite number of operators which will obtain expectation values proportional to the temperature to the appropriate power. 
The large-frequency behavior of various correlators is thus expected to receive contributions from an infinite number of such operators,
which appear in the corresponding OPE.
In other words, we expect generically that the true holographic background  
should contain additional fields with profiles that are needed to reproduce higher order terms in the OPE at large Euclidean frequencies.  Naively, one would expect that for real frequencies
far below the temperature, all such operators should become important in determining low energy quantities such as charge diffusion (where the OPE badly diverges).  However, the holographic model suggests that
high scaling dimension operators decouple rapidly if their OPE coefficient does not grow factorially.   
In that case, the diffusion constant and d.c.\@\xspace conductivity can be well described with only the lowest dimension operators.  If, on the other hand, the OPE coefficients grow rapidly, compensating
for the suppression factors found above, the holographic background can deviate considerably from the naive AdS$_4$-Schwarzschild form.
In fact, higher spin fields in the bulk (corresponding to higher spin CFT operators) 
can become important, spoiling the simple background-metric description. 
In such a situation, one would question not only the photon equation of motion \req{hol-gauge-lag} but also the boundary 
conditions used for the bulk modes at $u=1$.  Thus, a natural conjecture is that it is precisely for theories where the OPE coefficients
do not grow considerably that finite temperature can be modeled with a horizon.  In those theories, the leading correction to the low frequency conductivity should come from the lowest dimension operator,
as we have considered.



\subsection{Comparing holography with quantum Monte Carlo} \label{sec:comparing}

\begin{figure}   
\centering
\subfigure[]{\label{fig:qmc_sig1} \includegraphics[width=3.8in]{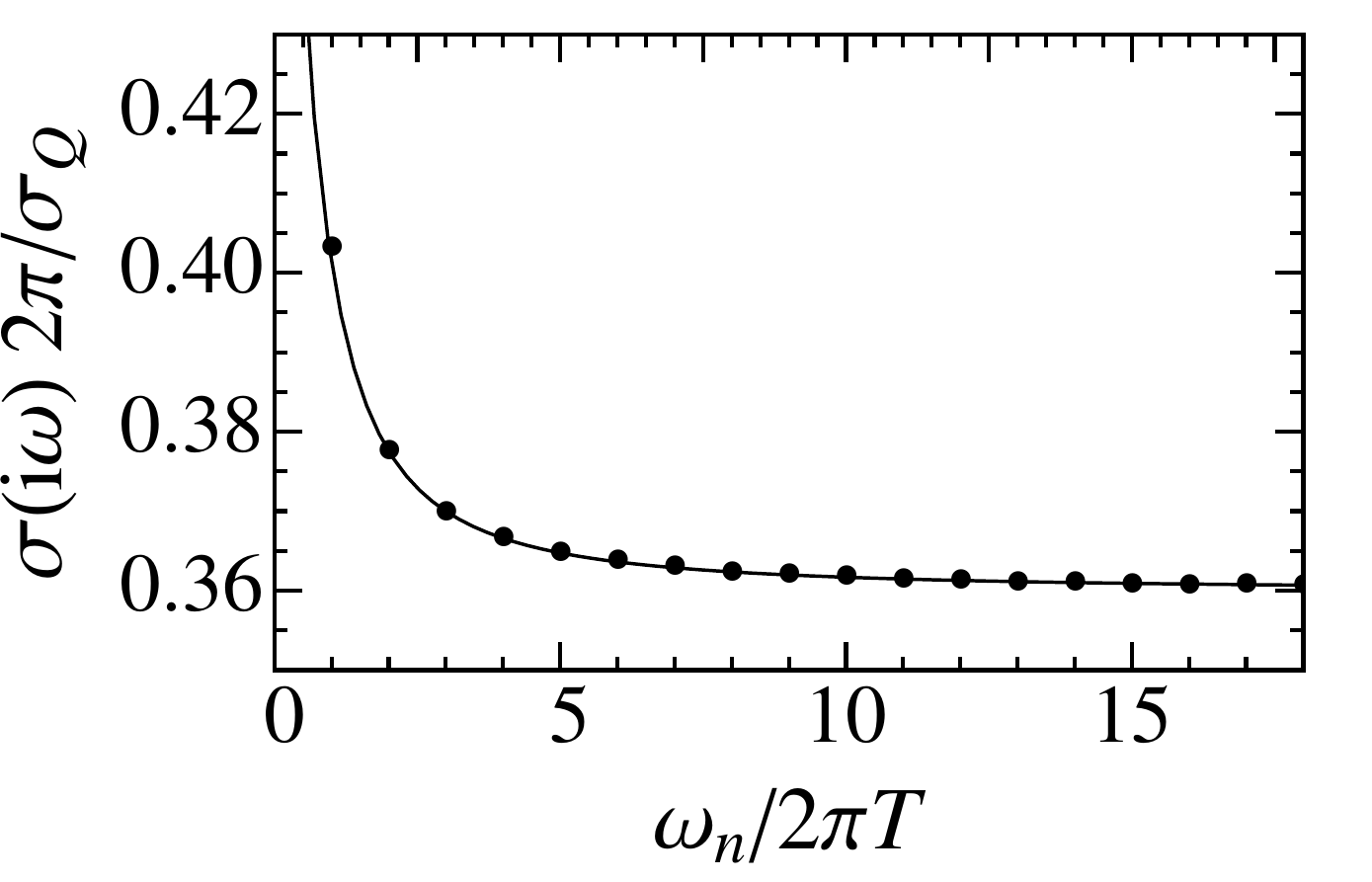}}       
\subfigure[]{\label{fig:qmc_sig-cont} \includegraphics[width=3.8in]{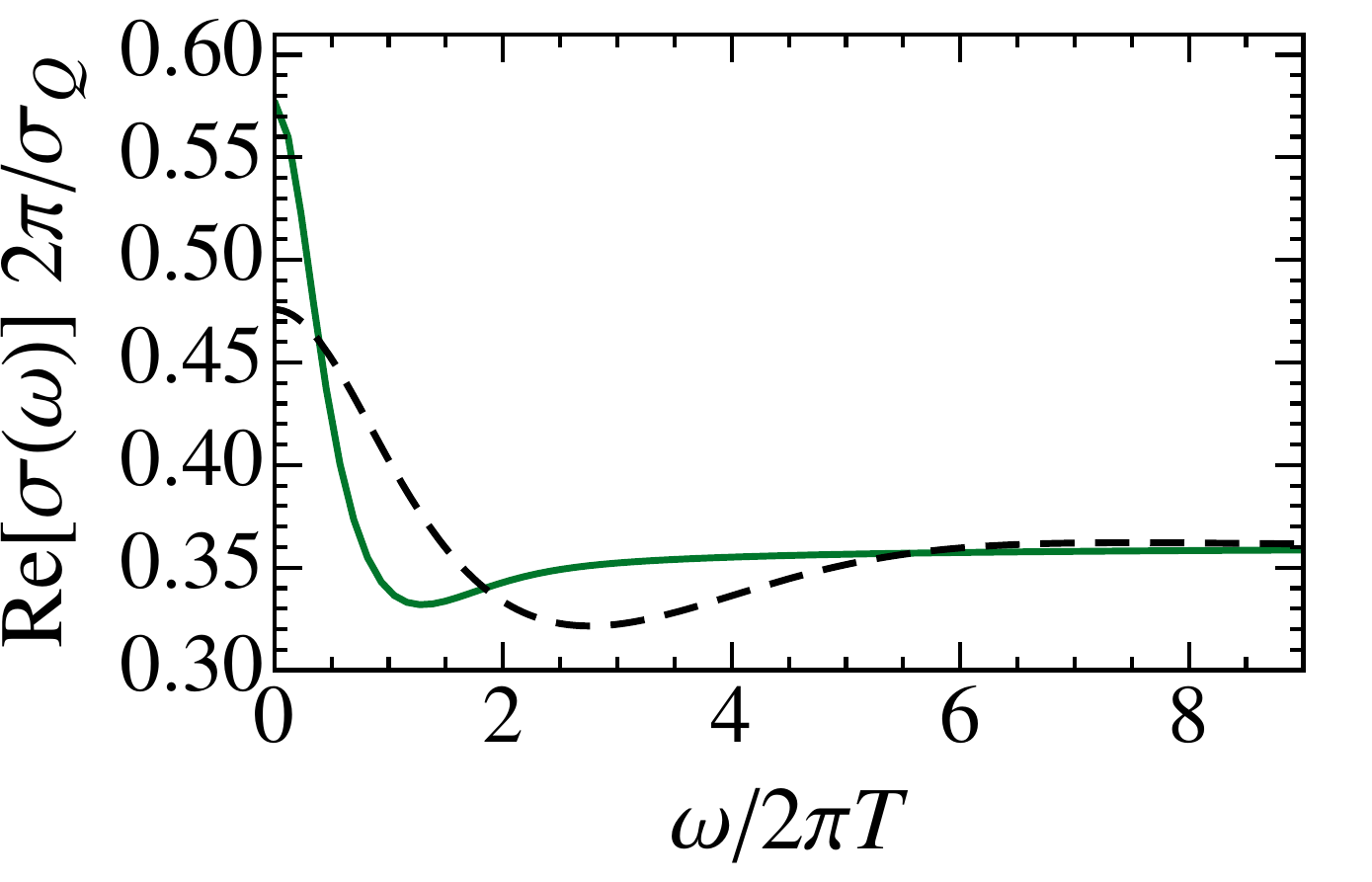}} 
\caption{\label{fig:qmc_sig} a) Holographic fit (line) to Quantum Monte Carlo data for the conductivity of a model in its O$(2)$ quantum 
critical regime (dots). The holographic parameters are:
$\Delta = 3/2, a\alpha = 0.6$. b) The corresponding conductivity on the real (Minkowski)
frequency axis (solid line). The dashed line corresponds to the holographic fit obtained in Ref.\ts\cite{krempa_nature},
where an ad hoc rescaling of temperature was needed.} 
\end{figure}    

We now solve the equation of motion for $A_y$, \req{A-EOM}, in order to study the full frequency dependence
of the conductivity, especially for real frequencies. 
We solve the differential equation numerically with in-falling boundary conditions at the 
horizon \cite{myers11}.
The solution can be obtained in the full complex plane of frequency. In particular, we can compare the holographic 
result with QMC data \cite{krempa_nature} for the O$(2)$ quantum critical theory, which is obtained for imaginary frequencies 
$\w_n\geq 2\pi T$, as shown in \rfig{qmc_sig1}. Most notably, we observe in \rfig{qmc_sig1} that the holographic result fits the
QMC data without the need of a temperature rescaling. A rescaling was needed previously \cite{krempa_nature,pollet} because
the holographic theory used then had the scaling dimension fixed to $\De=3$, \ie the dimension of the \enm tensor. 
In contrast, when the dimension is chosen to be that of the thermal operator $\De=\De_g=3-1/\nu\approx 1.5$, 
as expected from the OPE analysis above, a good fit results without the need for an ad hoc rescaling.
This fitting effectively determines the values of $b_1$ and $a \alpha$. We can now use these values to determine the conductivity
along the Minkowski frequency axis, and this leads to our main result in \rfig{qmc_sig-cont}.

\begin{figure}   
\centering
\includegraphics[width=3.9in]{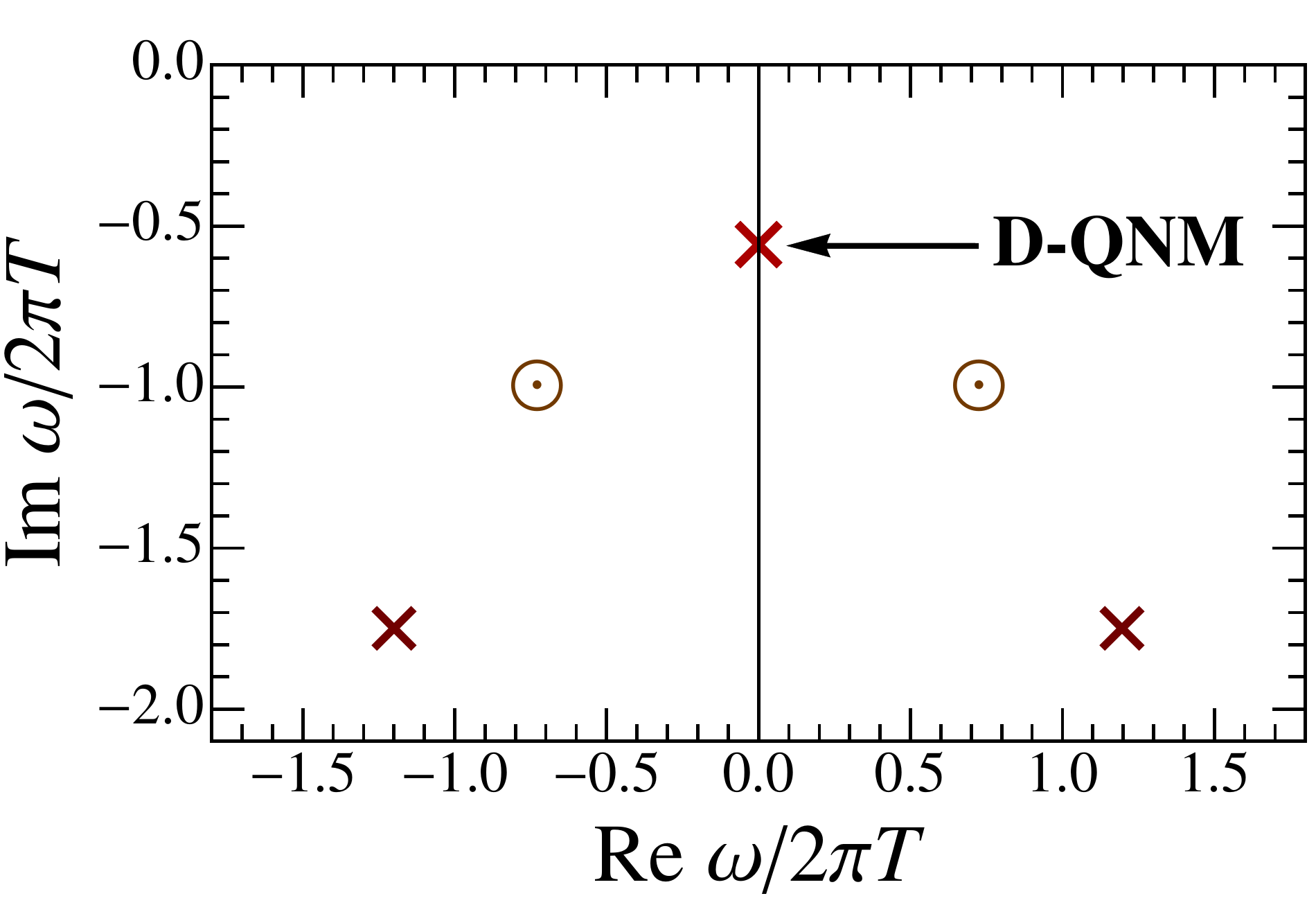}
\caption{\label{fig:qnm} The location of the small-frequency poles (crosses) and zeros (circles) of the holographic conductivity $\s(\w)$ in
the complex frequency plane. The parameters used are the same as those use to fit the O(2) QCP, see \rfig{qmc_sig}. The dominant,
purely damped pole is denoted by D-QNM, where QNM stands for quasinormal mode. } 
\end{figure}

We emphasize that certain qualitative features obtained using the previous holographic approach (which required rescaling)
remain unchanged with our new result, namely:
\begin{itemize} \itemsep0em
\item particle-like conductivity,
\item similar pole structure, \ie quasinormal spectrum (shown in \rfig{qnm}),
\item validity of sum rules \cite{sum-rules,ws}; see Section \ref{sec:sr}. 
\end{itemize}
The first two statements are related because a particle-like conductivity follows from the presence of a pole
on the negative imaginary-frequency axis, as shown in \rfig{qnm}. We note that such a purely damped pole for $\s(\w/T)$
was found in the O($N$) CFT at large-$N$ by including $1/N$ effects \cite{will-mit,ws}. 
In contrast, a vortex-like response would have a \emph{zero} on the imaginary axis;
see \rfig{sr} for two explicit examples.  
This purely damped pole dictates the ``topology'' of the full pole/zero spectrum 
as the poles and zeros appear in an alternating fashion. Mathematically, it follows because
the sign of the scalar coupling $\al$ dictates the presence of a particle-like ($\al>0$)
or vortex-like ($\al<0$) conductivity for any allowed $\De$. 

\section{Fermionic CFTs} 
\label{sec:fermions}

We briefly discuss extension to CFTs with Dirac fermions. A large class of such CFTs differ crucially from the O($N$) CFT by the 
absence of any scalar operator $\mathcal{O}$ in the $JJ$ OPE with scaling dimension $\Delta < 3$. Consequently, the leading term in the 
large $\omega$ dependence of the conductivity in Eqs.\ts\ref{eq:mainope} and (\ref{eq:sigmalargew}) is just given by that
from the OPE with the energy-momentum tensor. And such terms were implicitly accounted for in the previous holographic 
studies \cite{myers11,krempa_nature}.

The basic point is already evident from the CFT of  free (two-component) Dirac fermions.
The Lagrangian is
\begin{align}
\mc L=\bar\psi i\g_\mu\pd_\mu \psi\,, \label{eq:LDirac}
\end{align}
where $\g_\nu$ are the Euclidean gamma matrices $\g_\nu^\dag=\g_\nu$ satisfying the Clifford algebra $\{\g_\mu,\g_\nu\}=2\de_{\mu\nu}$. 
The conserved U(1) current is $J_\mu=\bar\psi\g_\mu\psi$. 
The integral expression for the finite-$T$ conductivity can be simply obtained:
\begin{align}
  \frac{\s(i\w_n)}{\sigma_Q} = \frac{1}{\w_n} T\sum_{\nu_n}\int \frac{d^2\vec k}{(2\pi)^2} \frac{1}{\e_k^2+\nu_n^2}
  \left[ \frac{4k_x^2+\w_n^2}{\e_k^2+(\nu_n+\w_n)^2} -\frac{4k_x^2}{\e_k^2+\nu_n^2}  \right]\,, \label{eq:sig-dirac}
\end{align} 
where $\nu_n=\pi T(2n+1)$ and $\e_k=k$.
This leads to the following high frequency behavior $\w_n\gg T$:  
\begin{align}
 \frac{\s(i\w_n)}{\sigma_Q} &= \frac{1}{16}-\frac{T}{2\pi\w_n}\sum_{m=1}\left(\frac{-T^2}{\w_n^2}\right)^m s_m \label{eq:dirac_asym} \\
  &= \frac{1}{16}+\frac{3\z(3)T^3}{\pi\w_n^3}- \frac{180\z(5)T^5}{\pi\w_n^5} +\mc O((T/\w_n)^7)\,, \label{eq:dirac_asym2}
\end{align} 
where $s_m=(2^{2m}-1)(2m)!\,\z(2m+1)$, and $\z$ is the Riemann zeta function. We refer the reader to 
Appendix~\ref{ap:dirac} for further details on the calculation.  

The most notable feature of  \req{dirac_asym2} is the absence of the $(T/\omega_n)^2$ term (found in \req{sigmaseries} for the 
$N=\infty$ O($N$) model), 
and the presence of a leading $(T/\omega_n)^3$ term. The latter corresponds to the term associated with the
energy-momentum tensor in \req{mainope}, and we show in Appendix~\ref{ap:dirac} that the
coefficient of $(T/\w_n)^3$ in \req{dirac_asym2}
is consistent with the value of the OPE coefficient $\mathcal{C}_T$. Such a $(T/\omega_n)^3$ term is clearly generic to all CFTs. 

The absence of a scalar operator with $\Delta < 3$ is also easily understood. A likely candidate for a scalar is $\bar\psi \psi$, but such 
a mass  term for Dirac fermions 
breaks both time-reversal and parity symmetries in 2+1 dimensions; this is the case even if such a mass term
acquires an expectation value only at finite temperature.
It is now also clear that such a scalar is also absent in interacting CFTs in which the Dirac fermions are coupled to gauge 
fields (such as QED3),
at least in the context of the $1/N_f$ expansion \cite{WenWu,Kaul2008,KlebanovSS}, where $N_f$ is the number of flavors of Dirac fermions. 
If the CFT has both Dirac fermions and elementary scalar fields $\phi_\alpha$
(as in the Gross-Neveu model), then in general an operator $\mathcal{O} \sim \phi_\alpha^2$ with $\Delta < 3$ 
will be generated at $T>0$ unless this is protected by additional symmetries, such as supersymmetry.

\section{Sum rules}\label{sec:sr}
The asymptotic behavior of the conductivity derived from the current-current OPE can be used to establish 
the finite-$T$ conductivity sum rules recently put forward \cite{sum-rules,ws,william2}:
\begin{align}
  \int_0^\infty d\w [\re\s(\w/T)-\s(\infty)]&=0\,, \, \label{eq:sr1} \\ 
  \int_0^\infty d\w \left[\re\left\{\frac{1}{\s(\w/T)}\right\}-\frac{1}{\s(\infty)}\right]&=0\,. \label{eq:sr2}
\end{align}
The second sum rule \cite{ws} is the S-dual or particle-vortex dual of the first one. 
An essential ingredient for the sum rules to be valid is that the integrand must be integrable.
Assuming this holds, one can extend the integration to be from $-\infty$ to $+\infty$, 
since in both cases the argument is even.
\req{sr1} can then be proven by performing a contour integration in the upper complex half-plane, where
$\s(z)$ is analytic by virtue of the retardedness of the current two-point function. A similar argument
holds for \req{sr2}, as we explain in Section~\ref{sec:dual-sr}. 

Our main objective is thus to show that the integrand decays 
sufficiently fast as $\w/T\ra\infty$. This is precisely the regime
where our OPE analysis applies. As we discussed above, see \req{mainope}, the operator with  
the smallest scaling dimension and finite thermal expectation value appearing in the current-current OPE 
dictates how fast $\re\s(\w/T)-\s(\infty)$ vanishes. Along the imaginary axis, the decay is 
$(T/\w_n)^{\De}$, where $\De$ is the dimension of 
the operator in question. Non-scalar operators, \ie with a finite spin $\ell>0$,
such as the energy-momentum tensor ($\ell=2$) cannot cause any problems at large frequencies because
their scaling dimension is guaranteed to be sufficiently large, being bounded from below by unitarity: 
$\De_{\ell>0}\geq \ell +1$, for CFTs in 2+1D. For instance,
the energy-momentum tensor saturates the $\ell=2$ bound yielding a $(T/\w_n)^3$ contribution to the  
conductivity on the imaginary axis. This term does not even contribute to $\re\s$ at real frequencies,  
which is of interest for the sum rule. In contrast, scalar operators ($\ell=0$) have the potential
of making the integrand of \req{sr1} non-integrable because of the weaker lower bound, $\De_{\ell=0}\geq (D-2)/2=1/2$.
However, in all the CFTs known to the authors, the scalars appearing the $JJ$ OPE have sufficiently high
scaling dimension to ensure that the sum rule \req{sr1} is well-defined. 
As it is difficult to make rigorous statements in general, we focus on the two families of CFTs discussed above. 

\subsection{O$(N)$ model}
For the O($N$) vector model, the leading operator in the $JJ$ OPE is the thermal operator $\mc O_g$ discussed above.
It has scaling dimension $\De_g=3-1/\nu$. We thus need $\De_g>1$, \ie $\nu>1/2$, for the sum rule to be well-defined. 
Now, for $N=2$, it is known from Monte Carlo that $\nu$ is slightly greater than $2/3$. Also, there is 
strong numerical and analytical evidence that $\nu$ increases with $N$, until it reaches the exact value
$\nu=1$ at $N=\infty$. We thus conclude that the conductivity of the O$(N)$ CFT decays sufficiently fast
for the sum rule to hold for all $N>1$. When $N=\infty$, the decay is $(T/\w)^2$ on the real axis, since $\De_g\big|_{N=\infty}=2$. 
In that case, the sum rule, \req{sr1}, was previously shown to hold by two of us \cite{ws}. 

\subsection{Fermionic CFTs} 
For the free Dirac CFT, we have shown that the leading operator that appears in the $JJ$ OPE is the energy-momentum
tensor, which has dimension $\De=3$, ruling out potentially dangerous scalars. An explicit analysis \cite{ws}
has indeed shown that the sum rule holds.
This is also the case for interacting CFTs in which $N_f$ Dirac fermions are coupled to gauge fields
(at least in the context of the $1/N_f$ expansion). These theories are thus expected to satisfy the sum rule \req{sr1}.

\subsection{Dual sum rule} \label{sec:dual-sr}
The dual sum rule, \req{sr2}, follows from the sum rule for $\s$ \req{sr1} for two reasons:
1) the large-frequency asymptotics of $1/\s$ are the same as those of $\s$ on the imaginary axis; 
2) $\s(z)$ has no zeros in the upper half-plane. The first point can be easily seen by inverting 
$\s(i\w_n)=\s_\infty + b_1(T/\w_n)^\De +\cdots$, and keeping the leading high-frequency term. It thus
shows that if $\re\s(\w/T)-\s(\infty)$ is integrable as $\w/T\ra\infty$, then 
$\re[1/\s(\w/T)]-[1/\s(\infty)]$ also is.
The second point follows from the analyticity of $\s(z)$ in the upper half-plane. It can be seen using the 
spectral representation of the current-current correlator.  

In Appendix~\ref{ap:sr}, we explicitly verify that the dual sum rule \req{sr2} is respected by both
the O$(N)$ model in the $N=\infty$ limit, and by the Dirac CFT. These constitute the first non-holographic
checks.

\section{Conclusions} 
Our paper has used the operator product expansion to obtain insight into the frequency dependence of the quantum-critical
conductivity near the superfluid-insulator transition in 2 spatial dimensions at non-zero temperatures; 
more generally, our results apply to conformal
field theories in 2+1 dimensions.

At frequencies $\omega \gg T$, we found that the conductivity had contributions $\sim (T/\omega)^{\Delta}$, where $\Delta$ is
the scaling dimension of any operator appearing in the OPE of two currents that acquires a non-zero expectation value at $T>0$. For the CFT describing the superfluid-insulator
transition, the smallest such $\Delta$ is that associated with the ``thermal'' operator $\phi_\alpha^2$ (where $\phi_1 + i \phi_2$ is the
complex superfluid order parameter), and this has scaling dimension $\Delta_g = 3- 1/\nu$, where $\nu$ is the correlation length
exponent. The next allowed operator is the energy-momentum tensor, which has $\Delta = 3$. The contribution of the energy-momentum
tensor is the leading term for CFTs which don't have allowed ``thermal'' operators, which includes wide classes of CFTs with Dirac fermions.

We computed the OPEs (and associated frequency dependence of the conductivity) of the $\phi_\alpha^2$ operator,
and of the energy-momentum tensor, for the O($N$) CFT using the vector $1/N$ expansion. These results, and prior computations for the O($N$)
CFT, were found to be in excellent agreement with quantum Monte Carlo simulations.

We then addressed the question of extending these $\omega \gg T$ results to smaller $\omega$. For all non-zero, Euclidean Matsubara frequencies,
the low frequency conductivity can be obtained in a controlled manner using the vector $1/N$ expansion. However, this expansion fails
for small real Minkowski frequencies \cite{damle}, and physically motivated resummations are required. For quantum systems with quasiparticle excitations,
the low frequency behavior is conventionally obtained by the Boltzmann equation. For strongly interacting CFTs {\em without\/} quasiparticles, we have
advocated \cite{krempa_nature} holographic methods. Here, we used the large $\omega$ behavior obtained from the 
OPE to determine the structure of the holographic
theory, and then solved the classical holographic theory to obtain the desired small $\omega$ dependence of the conductivity. 
In this holographic mapping, we truncated the OPE to the leading ``thermal'' operator, and presented evidence that the contributions
of high dimension operators can be suppressed even at low frequencies. 

Finally, we noted how conductivity the sum rules in Eqs.\ts(\ref{eq:sr1},\ref{eq:sr2}) can be established from 
information on the operator product expansion.

\section*{Acknowledgments} 
We thank J. Maldacena for pointing out in 2011 that the operator product expansion could be used
to determine the non-zero temperature conductivity at large frequencies.
We also thank R.~Myers for insightful discussions, and S.~Hartnoll for his comments on the manuscript.
W.W.-K.\ acknowledges a useful discussion with D.T.~Son about the connection between sum rules and OPEs.
E.K. was supported by DOE grant DEFG02-01ER-40676. 
E.S.S.\ acknowledges allocation of computing time at the Shared Hierarchical Academic Research Computing Network (SHARCNET:www.sharcnet.ca) 
and support by NSERC. S.S.\ was supported by the NSF under Grant DMR-1360789, the Templeton foundation, and MURI grant W911NF-14-1-0003 from ARO.
This research was supported in part by Perimeter Institute for Theoretical Physics (W.W.-K.\ and S.S.). 
W.W.-K.\ is grateful for the hospitality of the Max Planck Institute for the Physics of Complex Systems and l'\'Ecole de
Physique des Houches where parts of the work were completed.   
Research at Perimeter Institute is supported by the Government of Canada through Industry Canada 
and by the Province of Ontario through the Ministry of Research and Innovation.   \\

\appendix 

\section{Correlators of the energy-momentum tensor}
\label{app:t} 

Ref.\ts\cite{suvrat} obtained a number of results for the 3-point correlator between the energy-momentum tensor
and the conserved O($N$) current. This appendix will translate those results into the form required for the OPE in \req{mainope}.

\subsection{O$(N)$ model}

First, we consider the correlators of the O($N$) theory in \req{Z} at its critical point for $N=\infty$. The 2-point
correlator of the energy-momentum tensor is 
\bea
\frac{1}{N} \left\langle T_{\mu\nu} (\bk) T_{\rho \sigma} (-\bk) \right\rangle 
&=& \frac{k^3}{512} \left( \delta_{\mu\rho} \delta_{\nu \sigma} + \delta_{\nu\rho} \delta_{\mu \sigma} - \delta_{\mu\nu} \delta_{\rho\sigma} + \delta_{\mu\nu} \frac{k_\rho k_\sigma}{k^2} + \delta_{\rho \sigma} \frac{k_\mu k_\nu}{k^2} \right. \nn
&~& \left. - \delta_{\mu\rho} \frac{k_\nu k_\sigma}{k^2} - \delta_{\nu\rho} \frac{k_\mu k_\sigma}{k^2}- \delta_{\mu\sigma} \frac{k_\nu k_\rho}{k^2}- \delta_{\nu\sigma} \frac{k_\mu k_\rho}{k^2}
+ \frac{k_\mu k_\nu k_\rho k_\sigma}{k^4} \right). \label{eq:et}
\eea

For the 3-point $TJJ$ correlator, from the results of Ref.\ts\cite{suvrat} we obtain
\beq
\lim_{|\omega| \gg p} \left\langle J_x (\bomega) J_x (-\bomega+\bp) T_{\mu\nu} (-\bp) \right\rangle = \mbox{contact terms} + 
\frac{O_{\mu\nu} (\bp)}{\omega^2} + \dotsb, \label{ope1}
\eeq
where $\bomega = (\omega, 0, 0)$. Some non-zero values of $O_{\mu\nu}$ are
\begin{align}
 O_{\tau\tau} &= 0\,,&  O_{xx} &= |\omega_1|^3/64\,,&  O_{yy} &= - |\omega_1|^3/64\,,&  \mbox{for $\bp = (\omega_1, 0, 0);$} \nn 
 O_{\tau\tau} &= 0\,,&  O_{xx} &= 0\,,& O_{yy} &= 0\,,& \mbox{for $\bp = (0, p_x, 0);$} \nn 
 O_{\tau\tau} &= - |p_y|^3/64\,,& O_{xx} &= |p_y|^3/64\,,& O_{yy} &= 0\,,& \mbox{for $\bp = (0, 0, p_y);$} \nn
& &  O_{\tau x} &= - |p|^3/(64 \sqrt{2})\,, & & & \mbox{for $\bp = (p, p, 0);\;\,$} \label{ope4}  
\end{align}   
To convert this information into an OPE, we need the two-point correlation matrix of the diagonal components of $T_{\mu\nu}$
which we define as $C_{\{\mu\nu\}} (\bp) = \left\langle T_{\mu\mu} (\bp) T_{\nu\nu} (-\bp) \right\rangle$. 
From \req{et} we obtain 
\beq
C_{\{\mu\nu\}} (\bp) = \frac{N|p|^3}{512} \left( \begin{array}{ccc}
0 & 0 & 0 \\
0 & 1 & -1 \\
0 & -1 & 1 
\end{array} \right) \quad , \quad \mbox{for $\bp = (p,0,0)$}, \label{ope2}
\eeq
and similarly for other orientations.

Now we assume the OPE
\beq
\lim_{|\omega| \gg p} J_x (\bomega) J_x (-\bomega-\bp) = 
\sum_\mu B_\mu \frac{T_{\mu\mu} (\bp)}{\omega^2} + \dotsb \label{ope3}
\eeq
Then from Eqs.\ts(\ref{ope4},\ref{ope2},\ref{ope3}) we have the constraints 
\bea
\frac{N}{512} (B_x - B_y) &=& \frac{1}{64} \nn
\frac{N}{512} (B_\tau - B_y) &=& 0 \nn
\frac{N}{512} (B_\tau - B_x) &=& -\frac{1}{64}\,.
\eea
From the last constraint in Eq.~(\ref{ope4}) we have
\beq
\frac{N}{512 \sqrt{2}} \left( - B_x - B_\tau + 2 B_y \right) = - \frac{1}{64 \sqrt{2}}\,.
\eeq
So a consistent solution (up to the vanishing trace) is 
\beq 
B_\tau = 0  \quad , \quad
B_x = \frac{8}{N} \quad , \quad B_y = 0\,.
\eeq
So we have our main result for the OPE of the O($N$) model
\beq
\lim_{|\omega| \gg p} J_x (\bomega) J_x (-\bomega-\bp) = 
\frac{8}{N} \frac{T_{xx} (\bp)}{\omega^2} + \dotsb . \label{ope30}
\eeq
From \req{mainope}, and using $\gamma=-1/12$ \cite{suvrat}, this leads to the value of $\mathcal{C}_T$ in \req{CTval}.  

\subsection{Fermions}

Next, we consider a theory of 2-component Dirac fermions with $N_f$ flavors, each with the Lagrangian in \req{LDirac}.
The 2-point correlator of the energy-momentum tensor has the same form
as \req{et}
\bea
\frac{1}{N_f} \left\langle T_{\mu\nu} (\bk) T_{\rho \sigma} (-\bk) \right\rangle 
&=& \frac{k^3}{256} \left( \delta_{\mu\rho} \delta_{\nu \sigma} + \delta_{\nu\rho} \delta_{\mu \sigma} - \delta_{\mu\nu} \delta_{\rho\sigma} + \delta_{\mu\nu} \frac{k_\rho k_\sigma}{k^2} + \delta_{\rho \sigma} \frac{k_\mu k_\nu}{k^2} \right. \nn
&~& \left. - \delta_{\mu\rho} \frac{k_\nu k_\sigma}{k^2} - \delta_{\nu\rho} \frac{k_\mu k_\sigma}{k^2}- \delta_{\mu\sigma} \frac{k_\nu k_\rho}{k^2}- \delta_{\nu\sigma} \frac{k_\mu k_\rho}{k^2}
+ \frac{k_\mu k_\nu k_\rho k_\sigma}{k^4} \right)\,. \label{eq:ef}
\eea

For the 3-point $TJJ$ correlator, the results of Ref.\ts\cite{suvrat} take the form in Eq.~(\ref{ope1}) with the following
values of $O_{\mu\nu}$ 
\begin{align}
 O_{\tau\tau} &= 0\,,&  O_{xx} &= |\omega_1|^3/64\,,&  O_{yy} &= - |\omega_1|^3/64\,,&  \mbox{for $\bp = (\omega_1, 0, 0);$} \nn 
 O_{\tau\tau} &= |p_x|^3/64\,,& O_{xx} &= 0\,,& O_{yy} &= -|p_x|^3/64\,,& \mbox{for $\bp = (0, p_x, 0);$} \nn 
 O_{\tau\tau} &= 0\,,&  O_{xx} &= 0\,,& O_{yy} &= 0\,,& \mbox{for $\bp = (0,0,p_y);$} \nn 
& & O_{\tau x} &= - |p|^3/(32 \sqrt{2})\, , & & & \mbox{for $\bp = (p, p, 0)\,.\;$} \label{eq:ope4p}  
\end{align}   
Now the constraints are 
\bea
\frac{N_f}{256} (B_x - B_y) &=& \frac{1}{64} \nn
\frac{N_f}{256} (B_\tau - B_y) &=& \frac{1}{64} \nn
\frac{N_f}{256} (B_\tau - B_x) &=& 0 \,.
\eea
From the last constraint in \req{ope4p} we have    
\beq
\frac{N_f}{256 \sqrt{2}} \left( - B_x - B_\tau + 2 B_y \right) = - \frac{1}{32 \sqrt{2}}
\eeq
So a consistent solution (up to the trace) is 
\beq 
B_\tau = 0  \quad , \quad
B_x = 0 \quad , \quad B_y = - \frac{4}{N_f}
\eeq
Then we have the main result for the OPE of the fermion theory
\beq
\lim_{|\w| \gg p} J_x (\b\w) J_x (-\b\w-\b p) = -
\frac{4}{N_f} \frac{T_{yy} (\b p)}{\w^2} + \dotsb. \label{ope31}
\eeq
From \req{mainope}, and using $\gamma=1/12$ \cite{suvrat}, this leads to 
\beq
\mathcal{C}_T = \frac{2}{N_f}\,.
\label{eq:CTval2}
\eeq

\subsection{Holography}

Using a holographic theory with Einstein-Maxwell terms along with a coupling $\gamma$ to the Weyl tensor,  
the results of Ref.\ts\cite{suvrat} translate to the following correlators (up to an overall normalization dependent upon Newton's constant)
\bea
\lim_{|\omega| \gg p} \left\langle J_x (\bomega) J_x (-\bomega-\bp) \Bigl( T_{xx} (\bp) - T_{yy} (\bp) \Bigr) \right\rangle &=& 
\frac{|p|^{3/2}}{4\omega^2} \quad , \quad \bp = (p,0,0)  \nn
\lim_{|\omega| \gg p} \left\langle J_x (\bomega) J_x (-\bomega-\bp) \Bigl( T_{yy} (\bp) - T_{\tau\tau} (\bp) \Bigr) \right\rangle &=& 
-\frac{(1+12\gamma)|p|^{3/2}}{8\omega^2} \quad , \quad \bp = (0,p,0)  \nn
\lim_{|\omega| \gg p} \left\langle J_x (\bomega) J_x (-\bomega-\bp) \Bigl( T_{xx} (\bp) - T_{\tau\tau} (\bp) \Bigr) \right\rangle &=& 
\frac{(1-12\gamma)|p|^{3/2}}{8\omega^2} \quad , \quad \bp = (0,0,p)  
\eea
We note that the above results are entirely consistent with the O($N$) model ($N\ra\infty$) results for $\gamma=-1/12$, and with the free fermion results for $\gamma = 1/12$,
just as expected.
For a general CFT, proceeding as in the previous subsections, we obtain \req{mainope}.

\section{Correlators of the O($N$) model at $T=0$}
\label{app:t0} 

\subsection{Two-point function of $\mc O_g$}
The $T=0$ correlators of $Z$ in \req{Z1} have been evaluated at some length in Ref.\ts\cite{podolskyss},
including the two-point correlator of $\mathcal{O}_g$. We recall here the needed results. 

The computation proceeds by expanding about the large $N$ saddle point of \req{Z1} after setting $v=\infty$. 
We denote the saddle point value of $i \widetilde{\lambda}$ as $\sqrt{N} r$, and the fluctuation about the saddle point as $i \lambda$:
\beq
\widetilde{\lambda} = - i \sqrt{N} r + \lambda .
\eeq
The equation determining the value of $r$ is
\beq
\frac{1}{g} = \int_\bp \frac{1}{p^2 + r}. 
\label{eq:eqgr}
\eeq
The quantum critical point has $r=0$ at $T=0$, and so it is $g=g_c$ where
\beq
\frac{1}{g_c} = \int_\bp \frac{1}{p^2}.
\eeq
A standard $1/N$ expansion then yields the 2-point correlator of $\lambda$ as \cite{podolskyss}
\bea
G_{\lambda\lambda} (p) &=& 16 p 
- \frac{512}{N} \int_\bk \frac{1}{k |\bk-\bp|} + \frac{256p}{N} \int_\bk 
\frac{1}{(\bp \cdot \bk)} \left[ \frac{1}{|\bk-\bp|}  -  \frac{1}{ |\bk+\bp|} \right] + \frac{512}{N} \int_\bk \frac{(\bp \cdot \bk)}{p k^2 |\bp-\bk|} \nn
&=& 16p - \frac{128}{N \pi^2} \left( 2 \Lambda - p \right) + \frac{256p}{N \pi^2} \ln \left( \frac{\Lambda}{p} \right)
+ \frac{256p}{3 N \pi^2} \left( \ln \left( \frac{\Lambda}{p} \right) + \frac{1}{3} \right)\nn
&=& - \frac{256 \Lambda}{N \pi^2} + 16p \left[ 1 + \frac{64}{3 \pi^2 N} \left(\ln \left(\frac{\Lambda}{p}\right)  + \frac{11}{24} \right) \right];
 \label{gll}
\eea
the last line above corrects a typographical error in the last line of Eq.~(B14) of 
Ref.\ts\cite{podolskyss}. Here $\Lambda$ is a relativistic hard-momentum cutoff.
The scaling dimension of $\lambda$ is the same as that of $\phi_\alpha^2$, which is $3-1/\nu$, and so using Eqs.~(\ref{eq:otot},\ref{CT}) 
we verify that we have at order $1/N$
\beq
G_{\lambda\lambda} (p)  \sim \mbox{constant} + \frac{16}{C_\lambda^2} \, p^{3 - 2 /\nu}, \label{GClam}
\eeq
with the exponent $\nu$ given by
\beq
\nu = 1 - \frac{32}{3 \pi^2 N} + \mathcal{O}(1/N^2),
\eeq
and 
\beq
C_\lambda = \Lambda^{1-1/\nu} \left( 1 - \frac{44}{9 \pi^2 N} + \mathcal{O}(1/N^2) \right). \label{clambdaval}
\eeq

\subsection{Three-point function}
\begin{figure}   
\centering
\includegraphics[width=2.2in]{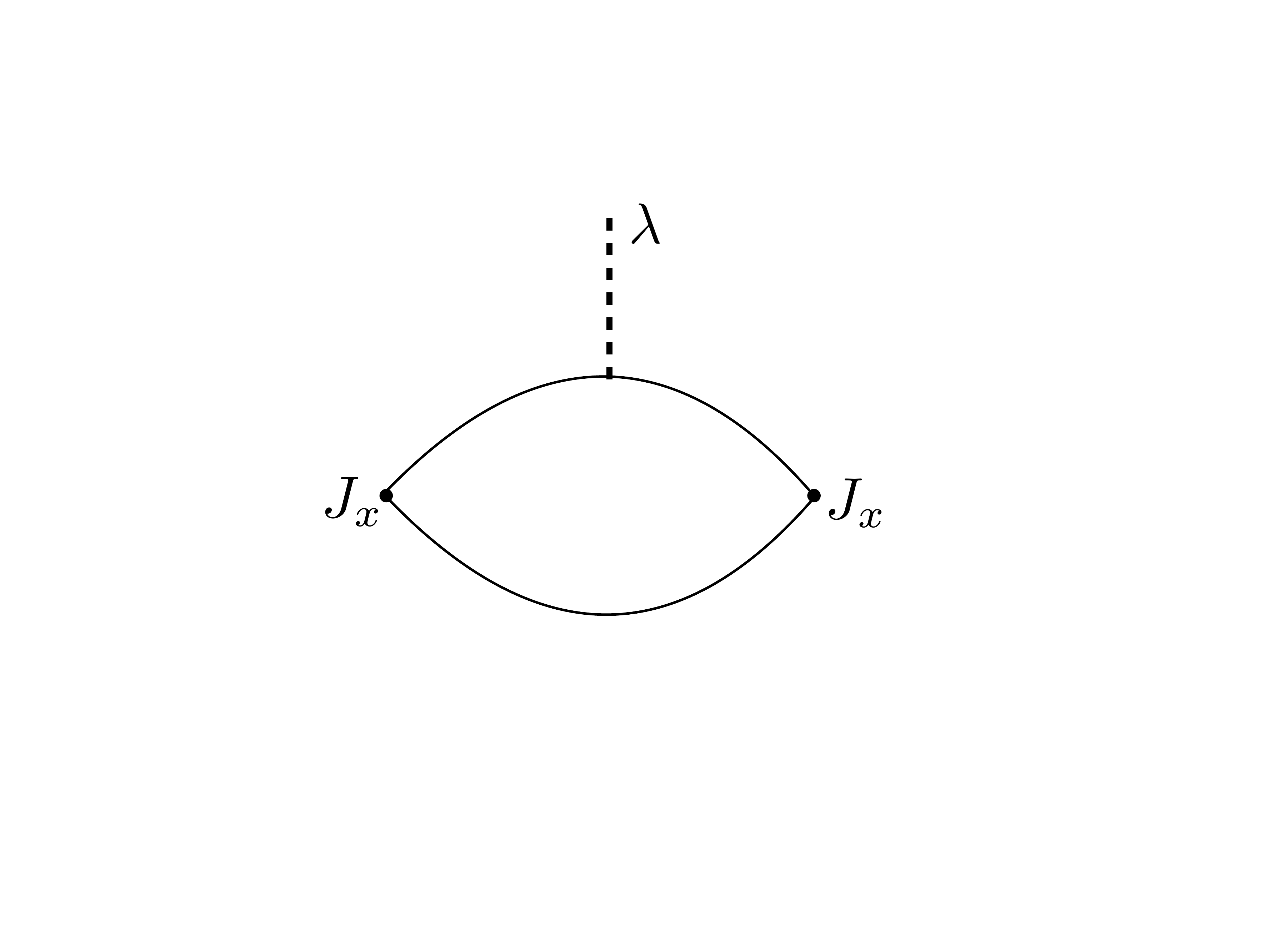}
\caption{\label{fig:Cg} 
Feynman graph for the computation of $\mathcal{C}_g$, \req{jjo}. The full line is the $\phi_\alpha$ propagator, and the dashed line
is the $\lambda$ propagator.} 
\end{figure}

To determine the OPE coefficient $\mathcal{C}_g$ in \req{mainope} we compute the associated 
3-point correlator, as in Eq.~(\ref{ope1}). At leading order in $1/N$, this is given by the Feynman graph in Fig.~\ref{fig:Cg}, and leads to
\bea
\left\langle J_x (\bomega) J_x (-\bomega+\bp) \mathcal{O}_g (-\bp) \right\rangle &=&
\frac{32p}{\sqrt{N}} 
\int \frac{d^2 \vec k}{4 \pi^2} \int_{-\infty}^\infty \frac{d \epsilon}{2 \pi} \frac{4 k_x^2}{(\epsilon^2 + k^2)^2 
  ((\epsilon+\omega)^2 + k^2)} \label{eq:jjo} \nn
&=& \frac{4p}{\sqrt{N} |\omega|}\,,
\eea
where we have retained only the leading term in the $p\ra0$ limit.   
Using \req{otot}, we then obtain \req{Cgval}.

\section{Correlators of the O($N$) model at $T>0$}
\label{app:gt0}

An extensive study of the $T>0$ correlators of the O($N$) CFT was provided in Ref.\ts\cite{CSY} using the $1/N$ expansion.
Here we present the extensions needed for our purposes.

The first step in the $1/N$ expansion is the determination of the saddle-point value of $i \widetilde{\lambda}$. Solving the extension
of \req{eqgr} at $T>0$ and $g=g_c$ now yields \cite{CSY} 
\beq
r = \Theta^2 T^2\,,
\label{eq:Theta}
\eeq
where $\Theta$ is given in \req{Thetaval}. 

For the computation of $\left\langle \mathcal{O}_g \right\rangle_T$ at $T>0$, we need the following polarization functions, defined in Ref.\ts\cite{CSY}, which 
determine the propagator of $\lambda$:
\bea
\Pi (q, \epsilon_n) &=& T \sum_{\omega_n} \int \frac{d^2 \vec k}{4 \pi^2} \frac{1}{(k^2 + \omega_n^2 + \Theta^2 T^2)((\vec k + \vec q)^2 + (\epsilon_n+ \omega_n)^2 + \Theta^2 T^2)} \nn
&=& \frac{1}{8 (q^2 + \epsilon_n^2)^{1/2}} + \frac{ (2 \epsilon_n^2 - q^2) \Theta^3 T^3}{(q^2 + \epsilon_n^2)^3} \frac{(1 - 6 \Xi)}{3 \pi}
+\mathcal{O} \left( \frac{T^5}{(q, \epsilon_n)^6} \right) \nn
\Pi (0,0) &=& \frac{\sqrt{5}}{8 \pi \Theta T} \nn
\Pi_2 (q, \epsilon_n) &=& T \sum_{\omega_n} \int \frac{d^2 \vec k}{4 \pi^2} \frac{1}{(k^2 + \omega_n^2 + \Theta^2 T^2)^2 ((\vec k + \vec q)^2 + (\epsilon_n+ \omega_n)^2 + \Theta^2 T^2)} \nn
&=& \Pi (0,0) \frac{ (q^2+ \epsilon_n^2) }{(q^2 + \epsilon_n^2)^2 + 4 \Theta^2 T^2 \epsilon_n^2}
+\mathcal{O} \left( \frac{T^3}{(q, \epsilon_n)^6} \right) \label{eq:csyres}
\eea
where 
\beq
\Xi = \frac{1}{\Theta^3} \int_\Theta^{\infty} dx \frac{x^2}{e^x - 1}.
\eeq
From these ingredients, a perturbative expansion from the $T>0$ version of the action in \req{Z1} yields \cite{CSY,podolskyss}
\beq
\left\langle \mathcal{O}_g \right\rangle_T = C_\lambda \sqrt{N} \left[\Theta^2 T^2 - \frac{2}{N} \frac{1}{\Pi (0,0)}\int \frac{d^2 q}{4 \pi^2} 
 T \sum_{\epsilon_n} \frac{\Pi_2 (q, \epsilon_n)}{\Pi (q, \epsilon_n)} + \mathcal{O}(1/N^2) \right] \label{og1}
\eeq
From \req{csyres} we can extract out the portion of integral which has a quadratic ultraviolet divergence
\bea
\frac{1}{\Pi (0,0)}  \int \frac{d^2 \vec q}{4 \pi^2}  T \sum_{\epsilon_n} \frac{\Pi_2 (q, \epsilon_n)}{\Pi (q, \epsilon_n)}
&=& \int \frac{d^2 \vec q}{4 \pi^2}  T \sum_{\epsilon_n}  \left[ \frac{\Pi_2 (q, \epsilon_n)}{\Pi (0,0) \Pi (q, \epsilon_n)}
- \frac{8}{(q^2 + \epsilon_n^2)^{1/2}} \right] \nn
&~&~~~~~~+ \int \frac{d^2 \vec q}{4 \pi^2}  T \sum_{\epsilon_n}  \frac{8}{(q^2 + \epsilon_n^2)^{1/2}} \label{eq:quadsub}
\eea
Examination of the subleading terms from \req{csyres} now shows that the first integral in \req{quadsub} only has a logarithmic
dependence upon the upper cutoff, and there is fortunately no $\Lambda T$ term --- such a term would violate scaling.
The second integral in \req{quadsub} is evaluated as
\bea
\int \frac{d^2\vec q}{4 \pi^2} T \sum_{\epsilon_n} \frac{8}{(q^2 + \epsilon_n^2)^{1/2}} &=& 
 \int \frac{d^2\vec q}{4 \pi^2} \int \frac{d \epsilon}{2 \pi} \frac{8}{(q^2 + \epsilon^2)^{1/2}} +
 2 \int \frac{d^2\vec q}{4 \pi^2} \int_q^\infty \frac{d \omega}{\pi} \frac{8}{\sqrt{\omega^2 - q^2}} \frac{1}{(e^{\omega/T} - 1)}
  \nn
&=& \int \frac{d^3 p}{8 \pi^3} \frac{8}{p} + \int_0^{\infty} \frac{d \omega}{\pi} \frac{8\omega}{ \pi (e^{\omega/T} - 1)}  \nn
&=& \frac{4 \Lambda^2}{\pi^2} + \frac{4T^2}{3}. \label{og2}
\eea
The $4 T^2/3$ term can also be obtained by zeta-function regularization in which we replace $\sqrt{q^2 + \epsilon_n^2}$ by 
$(q^2 + \epsilon_n^2)^s$ and analytically continue to $s=1/2$.
We numerically evaluated the first integral in \req{quadsub} by the methods of Ref.\ts\cite{CSY}, using a cutoff $\epsilon_n^2 + q^2 < \Lambda^2$,
and obtained
\beq
\int \frac{d^2\vec q}{4 \pi^2}  T \sum_{\epsilon_n}  \left[ \frac{\Pi_2 (q, \epsilon_n)}{\Pi (0,0) \Pi (q, \epsilon_n)}
- \frac{8}{(q^2 + \epsilon_n^2)^{1/2}} \right] = -\Theta^2 T^2 \left( \frac{16}{3 \pi^2} \ln \left(\frac{\Lambda}{T} \right) + 0.74145 \right)
\label{pisum}
\eeq
From Eqs.\ts(\ref{og1}), (\ref{og2}), and (\ref{pisum}) we obtain the needed expectation value
\beq
\left\langle \mathcal{O}_g \right\rangle_T - \left\langle \mathcal{O}_g \right\rangle_{T=0} = 
T^{3 - 1/\nu} \Lambda^{1/\nu-1} C_\lambda \sqrt{N} \Theta^2 \left[ 1 - \frac{1.3961}{N} + \mathcal{O}(1/N^2) \right]\,. 
\label{ogval}
\eeq
Using the value of $C_\lambda$ in Eq.\ts(\ref{clambdaval}) we see that Eq.\ts(\ref{ogval}) is independent of $\Lambda$ and universal;
it leads to \req{ogval2}. 

\subsection{Thermal average of $T_{xx}$}
Next, we turn to the determination of the expectation value of the energy-momentum tensor, $T_{\mu\nu}$. 
Specifically we focus on $\langle T_{xx}\rangle_T$, which gives the pressure of the CFT. The final results are Eqs.\ts\ref{eq:P-general}
and \ref{eq:tilde-c}. We begin by the computation in the 
$N=\infty$ limit, in which case the pressure is given by
the average of $\partial_x \phi_\alpha \partial_x \phi_\alpha$, and leads to 
\bea
\frac{\left\langle T_{xx} \right\rangle_T}{N} &=& T\sum_{\omega_n} \int \frac{d^2\vec k}{4 \pi^2} \frac{k_x^2 }{(\omega_n^2 + k^2 + r)} \nn
&=& \int \frac{d^2 \vec k}{4 \pi^2} \frac{k^2}{2} \left[ \frac{n_B(\sqrt{k^2 + r})}{\sqrt{k^2 + r}} + \frac{1}{2 \sqrt{k^2 + r}} \right] \nn
&=& \frac{1}{4 \pi} \int_{\sqrt{r}}^{\infty} d \varepsilon (\varepsilon^2 - r) n_B(\varepsilon) + \int \frac{d^2 \vec k}{4 \pi^2} \frac{k^2}{4 \sqrt{k^2 + r}}
\nn
&=& \frac{1}{4 \pi} \int_{\sqrt{r}}^{\infty} d \varepsilon (\varepsilon^2 - r) n_B(\varepsilon)  
+ \frac{r^{3/2}}{12 \pi} \label{eq:p_int1} \\
&=&  \frac{2\zeta(3)}{5\pi}T^3 \,, \label{eq:Txx1}
\eea
where $r=\Theta^2 T^2$ as specified in \req{Theta}, $n_B(\varepsilon)$ is the Bose function.  
We have used zeta function regularization in the last step, which is equivalent to subtracting the VEV, $\langle T_{xx}\rangle_{T=0}$. 
We now provide details on how to evaluate the integral in \req{p_int1} to obtain \req{Txx1}. 
Scaling out the temperature, the integral reduces to:
\begin{align} \label{eq:p_int2}
  \int_\Theta^\infty dz (z^2-\Theta^2)n_B(Tz) = \G(3)\Li_3(1/\phi^2) + 2\Theta\G(2)\Li_2(1/\phi^2)\,,
\end{align}
where $\G(z)$ is the gamma function, and $\Li_n(z)$ the polylogarithm. We recall that $\Theta=2\ln\phi$, 
where $\phi=(1+\sqrt{5})/2$
is the golden ratio. The values of the dilogarithm and trilogarithm evaluated at $1/\phi^2$ are known 
(see Ref.\ts\onlinecite{subir-polylog} and references therein):
\begin{align}
  \Li_2(1/\phi^2) &= \frac{\pi^2}{15} - (\ln\phi)^2\,; \\  
  \Li_3(1/\phi^2) &= \frac{4\zeta(3)}{5} -\frac{2\pi^2}{15}\ln\phi +\frac{2}{3}(\ln\phi)^3 \,.
\end{align}
Substituting these in \req{p_int2}, we obtain the final result \req{Txx1}.  

\subsubsection{Relating the pressure to the free energy}
The pressure of a CFT can also be determined from its free energy.
The free-energy density $\mc F=-(\ln\mc Z)/V$ of a CFT in $D$ 
spacetime dimensions, where
$V$ is the volume of the system and $\mc Z$ the partition function, is given by \cite{subir-polylog}
\begin{align} \label{eq:free-energ}
  \mc F =\mc F_{T=0} - \frac{\G(D/2)\zeta(D)}{\pi^{D/2}} \t c\, T^D\,.
\end{align}
The universal constant was found \cite{subir-polylog} to be $\t c=4N/5$ in the $N=\infty$ limit of 
the O$(N)$ CFT at $D=2+1$,
so that $\mc F - \mc F_{T=0}= -(2\zeta(3)/5\pi)NT^3$. (In contrast, $\t c=N$ for $N$ free scalars.)
We note that the absolute value of this quantity is precisely equal to the pressure found above.  
This is not a coincidence, given the relation between the densities of the free energy and 
the energy, $\langle T_{\tau\tau}\rangle_T$, of a CFT \cite{petkou99}: 
\begin{align}
  \langle T_{\tau\tau}\rangle_T - \langle T_{\tau\tau}\rangle_{T=0} &= (D-1) (\mc F-\mc F_{T=0})\,, \\
  &= -(D-1) \frac{\G(D/2)\zeta(D)}{\pi^{D/2}} \t c\, T^D\,.
\end{align}
Using the traceless of the energy-momentum tensor, we find that the pressure is exactly as found above, namely  
\begin{align}
  \langle T_{xx} \rangle_T -\langle T_{xx} \rangle_{T=0}= \frac{\zeta(3)}{2\pi}\t c\, T^3\,, \label{eq:P-general}
\end{align}
with $\t c=4N/5$ in the $N=\infty$ limit. In fact, the $1/N$ correction to $\t c$ is known \cite{CSY}
\begin{align}
  \t c = \frac{4N}{5} - 0.3344\,. \label{eq:tilde-c}
\end{align}
This leads to the refined estimate $H_{xx}\approx 0.24$ for the O$(2)$ CFT. 

\subsection{Conductivity} 
Finally, we determine the large frequency behavior of the conductivity by direct evaluation at $N=\infty$.
The conductivity at a Matsubara frequency $\omega_n$ is
\bea
\frac{\sigma (i\omega_n)}{\sigma_Q} &=& - \frac{4}{\omega_n} T \sum_{\epsilon_n} \int \frac{d^2\vec k}{4 \pi^2} \frac{k_x^2}{\epsilon_n^2 + k^2 + r} \left(
\frac{1}{(\epsilon_n + \omega_n)^2 + k^2 + r} - \frac{1}{\epsilon_n^2 + k^2 + r} \right) \nn
&=& - \frac{2}{\omega_n} \int \frac{d^2\vec k}{4 \pi^2} k^2 \left( \frac{1 + 2 n_B(\varepsilon_k)}{\varepsilon_k 
( \omega_n^2 + 4 \varepsilon_k^2)}
-  \frac{1}{4 \varepsilon_k^3} - \frac{[n_B(\varepsilon_k)]^2}{2 T \varepsilon_k^2}
-  \frac{(1 + \varepsilon_k/T) n_B(\varepsilon_k) }{2 \varepsilon_k^3} \right),
\eea
where $\varepsilon_k = \sqrt{k^2 + r}$. After a change of variables of integration we obtain our key result for the large $\omega_n$ expansion 
of the conductivity:
\bea
\frac{\sigma (i\omega_n)}{\sigma_Q} &=& - \frac{1}{\pi \omega_n} \int_{\sqrt{r}}^{\infty}  d \varepsilon \, \varepsilon (\varepsilon^2 - r)  \left( \frac{1 + 2 n_B(\varepsilon)}{\varepsilon
( \omega_n^2 + 4 \varepsilon^2)}
-  \frac{1}{4 \varepsilon^3} - \frac{[n_B(\varepsilon)]^2}{2 T \varepsilon^2}
-  \frac{(1 + \varepsilon/T) n_B(\varepsilon) }{2 \varepsilon^3} \right) \label{eq:rot-sig-int} \\
&=& \frac{1}{16} + \frac{1}{\omega_n} \left( \frac{\sqrt{r} - 2 T \ln (e^{\sqrt{r}/T} -1 )}{2 \pi} \right) + \frac{r}{4 \omega_n^2} \nn
&~&\qquad + 
\frac{1}{\omega_n^3} \left( - \frac{2 r^{3/2}}{3 \pi} 
- \frac{2}{\pi} \int_{\sqrt{r}}^{\infty} d\varepsilon \, (\varepsilon^2 - r) n_B(\varepsilon)
\right) + \mathcal{O} ( 1/\omega_n^{4}).
\label{eq:sigmaseries}
\eea
Note that for the value of $r$ in \req{Theta}, the coefficient of $1/\omega_n$ vanishes, as it must for 
agreement with \req{mainope}.
The  remaining terms in \req{sigmaseries} also agree precisely with \req{sigmaT} after insertions of the values
of the OPE coefficients and $T>0$ expectation values summarized in Section~\ref{sec:on}.

\section{Numerical Simulations}
\label{app:qmc}
We summarize some of the details of the numerical simulations along with the extrapolation procedures needed to analyze the results. 
Further details can be found in the supplementary material of Ref.\ts\onlinecite{krempa_nature}.

As described in the main text, the numerical simulations are performed using the Villain model~\cite{Villain} defined on
a $2+1$ dimensional discrete lattice of dimensions $L\times L\times L_\tau$ with $L_\tau\Delta\tau=\beta U$:~\cite{Cha91,sorensen92,Wallin94} 
\begin{equation}
  Z_{V}\approx 
{\sum_{\{\bf J\}}}'
\exp\left[-\frac{1}{K}
\sum_{(\tau,{\vec r})}\left( 
    \frac{1}{2}{\bf J}^2_{(\tau,{\vec r})}-\frac{\mu}{U} J^\tau_{(\tau,{\vec r})}   
    \right)\right] \ . 
\end{equation}
Here the sum, ${\sum_{\{\bf J\}}}'$, is over configurations with $\nabla\cdot {\bf J}=0$ and for the simulations we perform here $\mu=0$.
As pointed out above, apart from its simplicity, a significant advantage of this model is its explicit isotropy in space and time.
This isotropy is consistent with the fact that the dynamical critical exponent, defined through $\xi_\tau\sim\xi^z$, has the value $z=1$. 
When performing finite-size scaling
studies, simulations are therefore always performed with $L_\tau = c L$, with $c$ a constant close to 1. 
In our simulations,  typically more than $10^9$ Monte Carlo steps are performed for each simulation using very efficient {\it directed} Monte Carlo sampling~\cite{aleta,aletb}
allowing us to study systems with up to $320\times 320$ sites with $L_\tau=160$. For the Villain model the quantum critical point has been determined
with increasing precision~\cite{sorensen92,aleta,neuhaus,krempa_nature,pollet} and using histogram techniques we have determined it to be $K_c=0.3330671(5)$~\cite{krempa_nature}
in agreement with Ref.\ts\onlinecite{pollet}.

In order to compare to the results obtained using the holographic and field-theoretical analysis it is first necessary to extrapolate our
results to the thermodynamic limit, $L\to\infty$, while keeping $L_\tau$ constant. This was done using two different methods.
First by directly extrapolating results for several different lattice sizes assuming finite size corrections of the form
$e^{a L}/L^\alpha,$
Since the size of the system in the temporal direction is kept constant at $L_\tau$ it is natural to expect such an exponential dependence of
the finite-size corrections  and typically one finds $a\sim 1/L_\tau$.
Alternatively, one can perform simulations more or less directly in the thermodynamic limit by restricting the simulations to the zero
spatial winding sector~\cite{batrouni,pollet} for a single system with $L>L_\tau$.  Typically one uses $L=2 L_\tau$. Note that in this case
winding number fluxtuations still persist in the temporal direction.
If the latter procedure is used, results very close to the thermodynamic limit can be obtained in a single simulation since the
main effect of increasing the lattice size in the spatial direction is to suppress winding 
number fluctuations in the spatial direction. The results shown in Figs.~\ref{fig:OTV} and Fig.~\ref{fig:OTOTV} have been obtained in this way.

Somewhat surprisingly, it turns out that for the conductivity an additional $T\to 0$ ($L_\tau\to\infty$) extrapolation at fixed $\omega_n/T$ of the $L\to \infty$ data is
necessary in order to recover the true universal conductivity in the quantum critical regime. This second extrapolation of the conductivity data for the Villain model
was performed in Ref.~\ts\onlinecite{krempa_nature} with the results shown in Fig.~\ref{fig:Sigma_V_large_n}.
As described in Ref.~\ts\onlinecite{krempa_nature}, in order to perform this second $T\to 0$ extrapolation
of the numerical data for the conductivity we
assume corrections to the $T\to 0$ form
of the conductivity arise from from the leading irrelevant operator 
in the quantum critical regime with scaling dimension $w$~\cite{GZ,jzj,pollet}. 
In the presence of a single irrelevant operator we assume the general form:
\begin{equation}
\sigma(\omega_n/T,T)/\sigma_Q=\sigma^{T\to 0}(\omega_n/T)/\sigma_Q + f(\omega_n/T)(T/U)^w+g(\omega_n/T)(T/U)^{2w}+\dotsb,
\end{equation}
with $f$ and $g$ both scaling functions of argument $\omega_n/T =2\pi n$. 
Since $\omega_n/T\geq 2\pi$, it seems reasonable to expect
that to leading order $f(x)$ and $g(x)$ behave as $\sim x^w$.
Furthermore, for the Villain model we use the dimensionless inverse temperature $U/T =L_\tau\Delta\tau$ and dimensionless frequency
$\omega_n/U=2\pi n T/U=2\pi n/(L_\tau\Delta\tau)$.
It is therefore natural to state the above equation directly in terms of $\omega_n/U$ 
and we arrive at the following form:
\begin{equation}
\sigma(n,L_\tau)/\sigma_Q = \sigma^{T\to 0}(n)/\sigma_Q-a(\omega_n/U)^w+b(\omega_n/U)^{2w}+\dotsb,
\label{eq:fincond1}
\end{equation}
with $n$ the Matsubara index and $a,b$ dimensionless constants (independent of $\omega_n$) determined in the fit. 
Leaving $w$ a free parameter in our fits we find $w=0.877(2)$. This form is quite close to the one used in Ref.~\ts\onlinecite{pollet}.

A closely related form can be obtained by assuming that the presence of a finite $\omega_n$ will constrain the power-law $\omega_n^w$
associated with the irrelevant operator in the following manner:
\begin{equation}
\sigma(n,L_\tau)/\sigma_Q = \sigma^{T\to 0}(n)/\sigma_Q-c (\omega_n/U)^we^{-d (\omega_n/U)}.
\label{eq:fincond2}
\end{equation}
In the absence of more explicit analytical justification, both Eqs.\ts(\ref{eq:fincond1}) and (\ref{eq:fincond2}) may be seen as phenomenological and it would be reassuring if the final
results did not depend on details of these forms. Hence, as a consistency check, we have verified that the exponential form in
Eq.\ts(\ref{eq:fincond2}) yield almost identical results for the final $T\to 0$ extrapolated conductivity when compared to results obtained using Eq.\ts(\ref{eq:fincond1}).
In the case of Eq.\ts(\ref{eq:fincond2}), with $c,d$ fitted constants, we obtain good fits with $w=0.887(3)$ in good agreement with the result obtained for $w$ from Eq.\ts(\ref{eq:fincond1}).

\section{Dirac fermions}
\label{ap:dirac}

\subsection{Conductivity} 
We focus on the two-point function of the conserved U(1) current of the Dirac fermion CFT described by \req{LDirac}. 
To simplify the expression for the conductivity, \req{sig-dirac}, we 
perform the sum using the usual contour integration method to obtain:
\begin{align}
  \frac{\s(i\w_n)}{\sigma_Q} = \frac{1}{2\pi\w_n}\int_0^\infty d\e \left\{ \left[ \frac12 - \frac{2\e^2}{4\e^2+\w_n^2}\right][1-2n_F(\e)]
  +\frac{\e}{T} [n_F(\e)]^2e^{\e/T}\right\}\,, \label{eq:sig-dirac-int}
\end{align} 
where we have changed variables from $|\vec k|$ to $\e_k=\e$. $n_F(\e)=1/[1+\exp(\e/T)]$ is the Fermi-Dirac distribution. 
Some of terms can be integrated to yield the exact result: 
\begin{align}
  \frac{\s(i\w_n)}{\sigma_Q} =\frac{1}{16}+\frac{\ln 2}{2\pi\w_n} - \frac{1}{\w_n}\int_0^\infty \frac{d\e}{\pi}
  \left[ \frac12 - \frac{2\e^2}{4\e^2+\w_n^2}\right] n_F(\e)\, .
\end{align}
To obtain the asymptotic expansion for $\s(i\w_n)$ valid at large frequencies $\w_n\gg T$, we can now Taylor
expand the integrand in powers of $1/\w_n$. This gives our main result for the asymptotic behavior
of the Dirac fermion conductivity, valid for $\w_n\gg T$: 
\begin{align}
  \frac{\s(i\w_n)}{\sigma_Q}  &= \frac{1}{16}-\frac{T}{2\pi\w_n}\sum_{m=1}\left(\frac{-T^2}{\w_n^2}\right)^m(2^{2m}-1)(2m)!\,\z(2m+1) \label{eq:dirac_exp} \\
  &= \frac{1}{16}+\frac{3\z(3)T^3}{\pi\w_n^3}- \frac{180\z(5)T^5}{\pi\w_n^5} + \frac{22680\z(7)T^7}{\pi\w_n^7} +\mc O((T/\w_n)^9)
\end{align}
where $\z(z)$ is the Riemann zeta function: $\z(3)\approx 1.202$, etc. We have used the following result
\begin{align}
  \int_0^\infty d\e\, \e^p n_F(\e)= T^{p+1}(1-2^{-p})\G(p+1)\z(p+1)\,,
\end{align}
where $\G(z)$ is the Gamma function.  The coefficient of the $(T/\omega_n)^3$ term agrees with that in \req{sigmaT}
upon using the value of $\mathcal{C}_T$ in \req{CTval2}, the value $\gamma = 1/12$ \cite{suvrat}, and the value of $H_{xx}$ in \req{Hvalf}.

\subsection{Thermal average of $T_{\mu\nu}$} 
The energy-momentum tensor for the free Dirac fermion CFT reads: 
\begin{align}
  T_{\mu\nu}(x)=\frac{1}{4}(\bar\psi i\g_\nu\pd_\mu\psi-\pd_\mu\bar\psi i\g_\nu\psi) + (\mu\leftrightarrow\nu)\,,
\end{align}
where $\g_\nu$ are the Euclidean gamma matrices $\g_\nu^\dag=\g_\nu$ satisfying the Clifford algebra $\{\g_\mu,\g_\nu\}=2\de_{\mu\nu}$.
We Fourier transform to energy-momentum space, using $\psi(\bx)=\int_\bk \psi_\bk e^{i\bk\cdot \bx}$
and $\bar\psi(\bx)=\int_\bk \bar\psi_\bk e^{-i\bk\cdot \bx}$, where $\int_\bk=\int d^3k/(2\pi)^3$,
which becomes $T\sum_{\nu_n}\int d^2\vec k/(2\pi)^2$ at finite temperature. We get:
\begin{align}
  T_{\mu\nu}(\bp)=\int d^3x\, T_{\mu\nu}(\bx) e^{-i\bp\cdot \bx} =-\frac14 \int_\bk\bar\psi_\bk[\g_\nu (2k_\mu+p_\mu)+\g_\mu (2k_\nu+p_\nu)]\psi_{\bk+\bp}\,.
\end{align}
We now take the expectation value, for which we will need the fermion two-point function: 
\begin{align}
  \ang{\bar\psi_\bk \g_\mu \psi_{\bk'}}=\de^{(3)}(\bk-\bk')\frac{2k_\mu}{k^2}\,,
\end{align}
where the factor of 2 comes from the trace $\tr\tfrac12\{\g_\mu,\g_\nu\}=2\de_{\mu\nu}$.
This expression is consistent with the real space correlator given in Ref.\ts\onlinecite{osborn}, 
 $\ang{\bar\psi(\bx)\g_\mu\psi(0)}=i x_\mu/(2\pi x^3)$.
We thus get
\begin{align}
  \ang{T_{\mu\nu}(\bp)}=-2\de^{(3)}(\bp)\int_\bk \frac{k_\mu k_\nu}{k^2} .\label{eq:T_vac_vev}
\end{align}
The integral is ultraviolet divergent. However, we are interested in the thermal
expectation value from which \req{T_vac_vev} has been subtracted: $\ang{T_{\mu\nu}(\bp)}_T-\ang{T_{\mu\nu}(\bp)}_{T=0}$.
This is finite and can be readily evaluated:
\begin{align}
  \ang{T_{yy}(\bp)}_T-\ang{T_{yy}(\bp)}_{T=0} &= \de^{(3)}(\bp)\int_0^\infty \frac{d\e}{2\pi} \e^2 n_F(\e) \\
  &= \de^{(3)}(\bp) \frac{3\z(3) T^3}{4\pi},
\end{align}
which yields
\beq
H_{yy}= H_{xx} = \frac{3 N_f \zeta(3) }{4 \pi}\,,
\label{eq:Hvalf}
\eeq
with $N_f$ flavors.   

\section{Dual sum rule} \label{ap:sr}
We show that the dual sum rule \req{sr2} is respected by the conductivities
of both the O$(N)$ model in the $N=\infty$ limit and the free Dirac CFT. These constitute the first 
explicit CFT checks beyond holography \cite{ws}. In both cases we must resort to numerical integration
to explicitly verify the sum rules.

\begin{figure}   
\centering
\includegraphics[scale=.43]{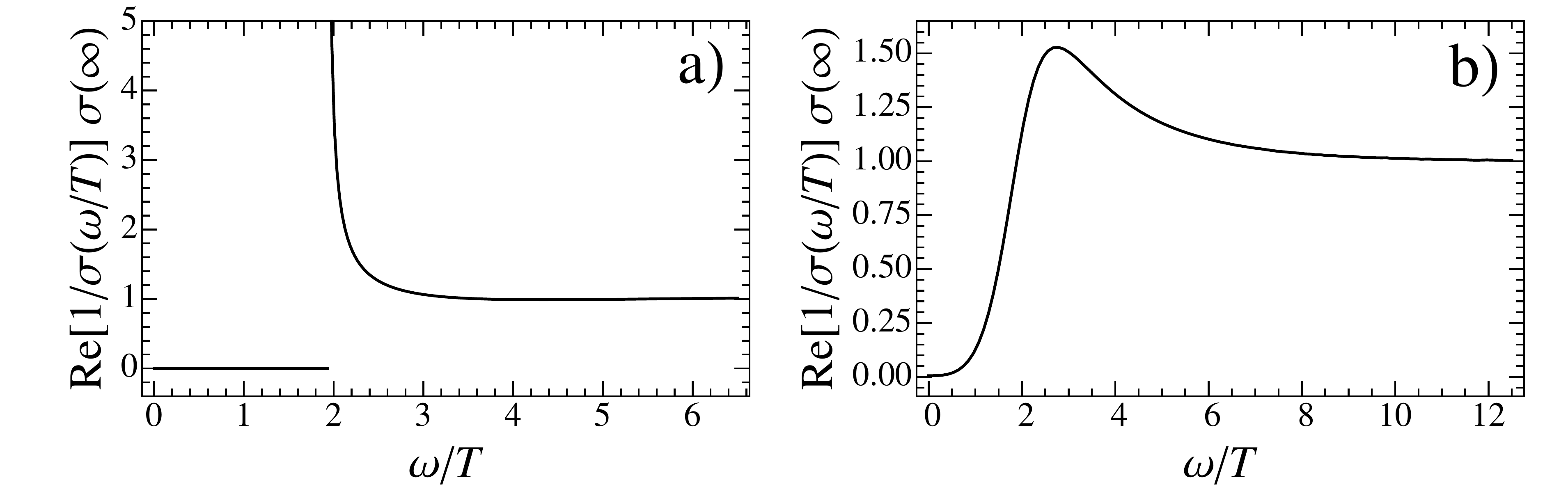}
\caption{\label{fig:sr} Real part of the dual conductivity, $\re[1/\s(\w/T)]$, for a) the O$(N)$ CFT
in the $N=\infty$ limit, b) the free Dirac CFT. Both constitute examples of vortex-like responses;
they respect the dual sum rule \req{sr2}. } 
\end{figure}

The conductivity of the O$(N)$ model in the $N=\infty$ limit is given by \req{rot-sig-int} for imaginary frequencies.  
In order to study the sum rule, we must analytically continue the expression to real frequencies $i\w_n\ra \w+i0^+$.
The resulting real part of the inverse conductivity is shown in \rfig{sr}a. Since $\s$ is particle-like \cite{ws},
$1/\s$ is vortex-like.
In fact, we find that a zero appears directly at the origin, $1/\s(0)=0$. This is as expected since the
direct conductivity $\s$ has a pole at $\w=0$ (leading to a delta-function in $\re\s$). 
At finite and small frequencies, 
a spectral gap naturally appears for $\re[1/\s]$ just as for $\re\s$. It is generated by the thermal mass $r^{1/2}=\Theta T$, \req{Theta}.
The numerical integration needed to establish \req{sr2} is complicated by the strong divergence of $\re[1/\s]$ seen at $\w=2r^{1/2}$:
\begin{align}
 \sim \frac{\Theta(\w-2r^{1/2})}{(\w-2r^{1/2})\left\{\ln[r^{1/2}/(\w-2r^{1/2})] \right\}^2}\,,
\end{align}
which is integrable, as it must be for the sum rule to hold. This divergence stems from the zero of the
conductivity, \ie a vanishing of both the real and imaginary parts, at $\w=2r^{1/2}$. This fact was uncovered in
Ref.\ts\onlinecite{ws}, where it was however erroneously concluded that the dual sum rule is not respected 
at $N=\infty$.
Here, we have carefully evaluated the integral, after having analytically computed the contribution near $\w=2r^{1/2}$,
and found that \req{sr2} holds. This is not surprising in light of the general arguments given in Section~\ref{sec:sr}.

The conductivity of the Dirac CFT is given by \req{sig-dirac-int}. 
The behavior of the inverse conductivity $1/\s$ is shown
for real frequencies in \rfig{sr}b. Just as for the $O(N)$ model discussed above, we find that it is vortex-like,
and vanishes at zero frequency: $1/\s(0)=0$. The numerical integration can be performed without difficulties to confirm the validity  
of the sum rule \req{sr2}.  

%

\end{document}